\documentclass[aps,prb,twocolumn,longbibliography,groupedaddress]{revtex4-2}

\usepackage{graphicx} 
\usepackage{graphics} 

\begin{document}


\title{Magnetoelectric effect in the mixed valence polyoxovanadate cage V$_{12}$}


\author{Piotr Kozłowski}
\email[]{kozl@amu.edu.pl}
\affiliation{Institute of Spintronics and Quantum Information, Faculty of Physics and Astronomy, Adam Mi\-ckie\-wicz University, Poznań, ul. Uniwersytetu Poznańskiego 2, 61-614 Poznań, Poland}


\date{\today}

\begin{abstract}
Development of spintronic and quantum computing devices increases demand for efficient, energy saving method of spin manipulation at molecular scale. Polyoxovanadate molecular magnets being susceptible to both electric and magnetic fields may serve here as a good base material. In this paper  two isostructural anions [V$_{12}$As$_8$O$_{40}$(HCO$_2$)]$^{n-}$ (with $n=3,5$) featuring two different mixed-valence states with itinerant and localized valence electrons are studied. The impact of the electric field on their magnetic properties is investigated by means of two complementary methods informed by magnetic measurements: effective Hamiltonian calculations and  density functional theory. It is demonstrated that the magnetoelectric effect in these molecules is induced mostly by relocation of itinerant electrons, is highly anisotropic, depends on the valence state and can be detected even at room temperature.
These findings can pave the way to practical applications in which an electric field control over spin state is required.

\end{abstract}


\maketitle

\section{Introduction}
Molecular magnets have been intensively investigated in the hope to engineer a quantum computer \cite{Leuenberger2001,Lehmann2007,timco2009,Chiesa2024} or an efficient information storage device \cite{Mannini2009}, and their role in the new field of spintronics \cite{hill2003,Bogani2008,urdampilleta2011,Vincent2012} is constantly increasing. 
These materials are also very interesting from a purely scientific point of view. Because of a well-defined structure, a small size and  a lack of intermolecular interactions between magnetic molecules various quantum phenomena, such as for example a discrete energy spectrum \cite{Baker2012b}, quantum tunneling of magnetization \cite{Thomas1996,Mannini2010}, Rabi oscillations \cite{Bertaina2008}, or quantum entanglement in its relation to magnetic frustration \cite{Kozlowski2015} can be precisely detected and analyzed.

One of the important aspects in view of envisaged technological applications is the manipulation method by which a magnetic state of a molecule can be changed.  
The most common and direct way of exerting such a control is the use of the magnetic field and then some suitable electron spin resonance \cite{Ardavan2007,Moro2014}. However this approach has some shortcomings. The spin manipulation must be done on very short spacial and temporal scales and a local application of fast changing magnetic field is problematic. 
Another way is the use of electric current to switch the magnetic state of the molecule \cite{Misiorny2007,Misiorny2010,Misiorny2013}, but such an approach may appear to be energy consuming.

An interesting alternative for spin manipulation in molecular magnets can be an application of the electric field. The main advantage of this method is a short manipulation time and no dissipation of energy, as no current is involved. Moreover, application of a scanning tunneling microscope (STM) tip enables generation of very strong fields in a small region, which can be rapidly changed by applying modulated voltage. Such a local control is most suitable for quantum information applications in which molecular magnets are considered to play a role of quibits \cite{Ferrando-Soria2016} or quidits \cite{Chicco2024} (for a recent review see \cite{Chiesa2024}) and in which quantum error correction can be applied \cite{Chiesa2020,Lim2025}. Electric field generated by STM tip can be used both to switch the spin \cite{Huang2025} and to gate two spins for instance in one polyoxometalic molecule by manipulating redox potential \cite{Lehmann2007} and giving rise to electrically controlled quantum gate.

Besides an STM tip also other methods are considered for electric manipulation of spin, such as for instance microwave cavity \cite{Trif2010}. For smaller fields and macroscopic samples a number of electric field based techniques has been used to induce and detect spin manipulation (magnetoelectric effect): electron paramagnetic resonance (EPR) with static electric field \cite{Boudalis2018}, a time resolved EPR experiment with electric pulses \cite{Liu2019,Robert2019}, a standard SQUID experiment in a static electric field \cite{Wang2023}, continues wave EPR with modulated electric field \cite{Kintzel2021,Cini2025}, low temperature susceptibility measurements in modulated electric field \cite{Lewkowitz2023} or  magneto-far-IR spectroscopy \cite{LeMardele2025}.

The influence of the electric field on the magnetic state of a molecule, which most of the time is due to the interaction with the valence electrons, is indirect and caused by modification of the molecular orbitals or/and spin-orbit interaction. It has been concluded \cite{Trif2008,Trif2010} on the basis of symmetry analysis that spin-electric coupling can appear in some magnetic molecules with permanent electric dipoles at bonds connecting various magnetic centers in a molecule. 
The effect of electric field can be then modeled by changing exchange interactions \cite{Trif2010}, but the strength of this effect cannot be assessed by symmetry analysis alone. For this purpose, it is necessary to perform for example DFT calculations \cite{Islam2010}.
A similar theoretical analysis has been performed for the time modulated electric field acting on a magnetic molecule \cite{Troiani2019}.
Another type of spin-electric coupling can be found in molecules which have structure related built-in electric dipole. In such compounds electric field induces structure change which in turn influences local magnetic anisotropy and thus the spin \cite{Liu2021,Vaganov2025}. It has also been demonstrated that in frustrated molecular triangles the spin-electric effect is caused by spin induced charge redistribution characteristic for chiral states \cite{Nossa2023}.

A different mechanism of interaction of a magnetic molecule with the electric field can be expected in systems with itinerant, that is delocalized valence electrons. Such situation is often encountered in the sufficiently symmetric mixed-valence polyoxovanadates (POV) \cite{Muller1991,Muller1997,Keene2012}. Here it is supposed that the most important effect is due to the displacement of itinerant electrons induced by the electric field. As a result, some electrons may  overcome Coulomb repulsion and  develop exchange interactions with each other. This, in turn, will lead to a modification of a molecular quantum state. It can be expected that in such molecules the spin-electric effect should be stronger than the one present in molecular magnets with only localized valence electrons. 
Moreover, POVs are of interest also because they can be easily functionalized by various ligands \cite{Monakhov2015b}  or by guests encapsulated in the host systems \cite{Monakhov2015,Kozlowski2017,Notario2018}  giving rise to a variety of properties which can be tuned to obtain desired effects. 

The theoretical estimation of the spin-electric effect in mixed-valence POVs has been limited only to molecules containing two unpaired electrons 
\cite{Cardona-Serra2013,Cardona-Serra2015,Palii2017}, or to dimers containing one V$^{3+}$ and one V$^{2+}$ ion \cite{Bosch-Serrano2012, Bosch-Serrano2012a}.
The experiments exploiting magnetoelectric effect in mixed-valence molecular magnets are very few \cite{Lehmann2007}.

The main goal of this paper is to investigate the magneto-electric coupling in the more general mixed-valence POV molecules containing both localized and itinerant electrons. To this aim two POV spherical isostructural molecules with different valence (6 and 8 unpaired electrons) featuring both itinerant and localized electrons are studied by means of two complementary theoretical methods: density functional theory (DFT) and effective Hamiltonian calculations. 
The parameters of the effective Hamiltonian are obtained by fitting the experimental magnetic data supported by the DFT calculations. The theoretical results obtained in this paper can be in principle verified experimentally by means of STM measurements.

The paper is organized as follows. In section II the molecules and the theoretical models are introduced. Determination of model parameters by fitting the magnetic measurements and DFT calculations is described in section III. In section IV the magnetic properties of both molecules for zero electric field are determined in details. Section V describes the influence of electric field and is followed by conclusions.

\section{V$_{12}$ molecules and models}
\begin{figure}[h]
	\centering	
	\includegraphics[height=6cm]{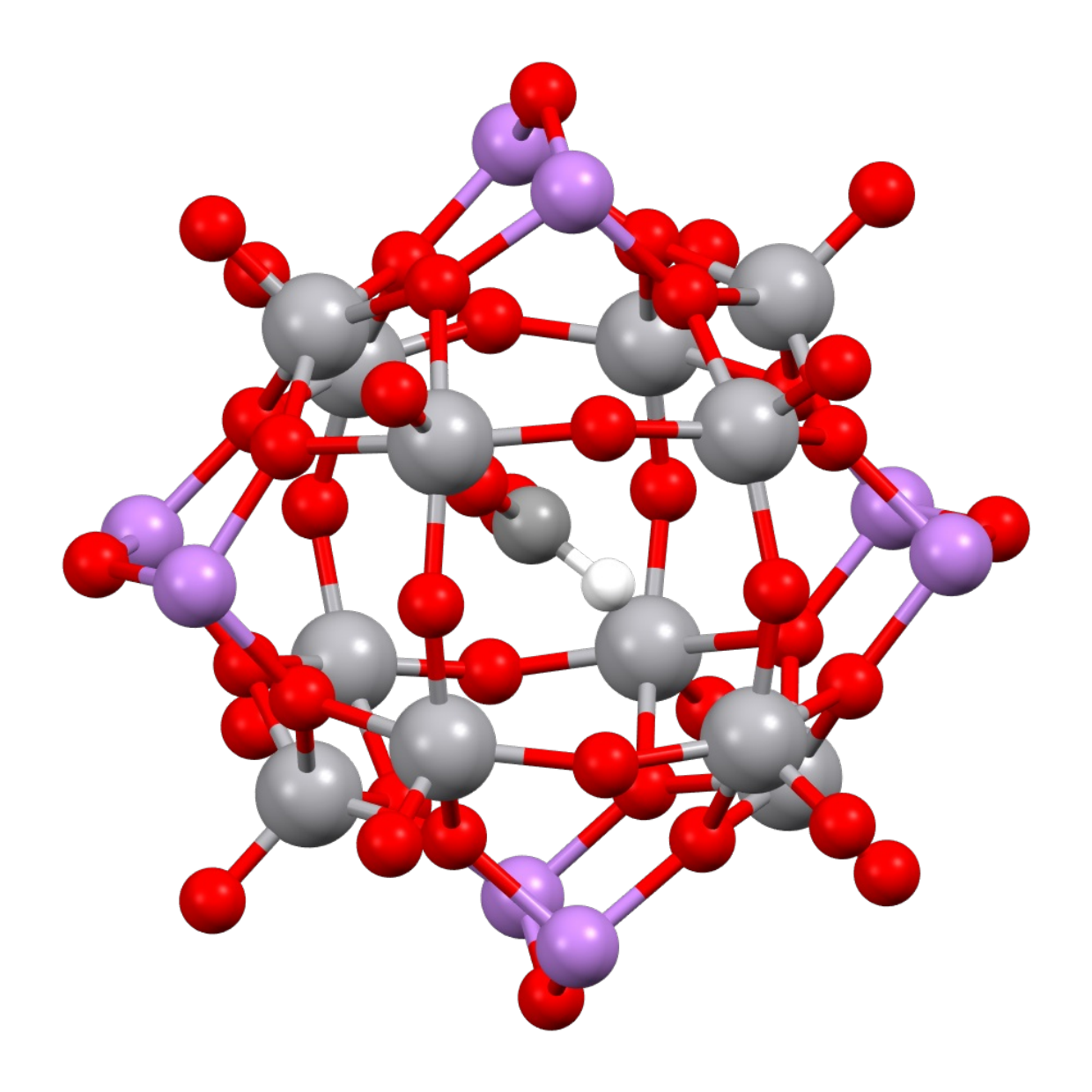}\\
	\includegraphics[height=4.5cm]{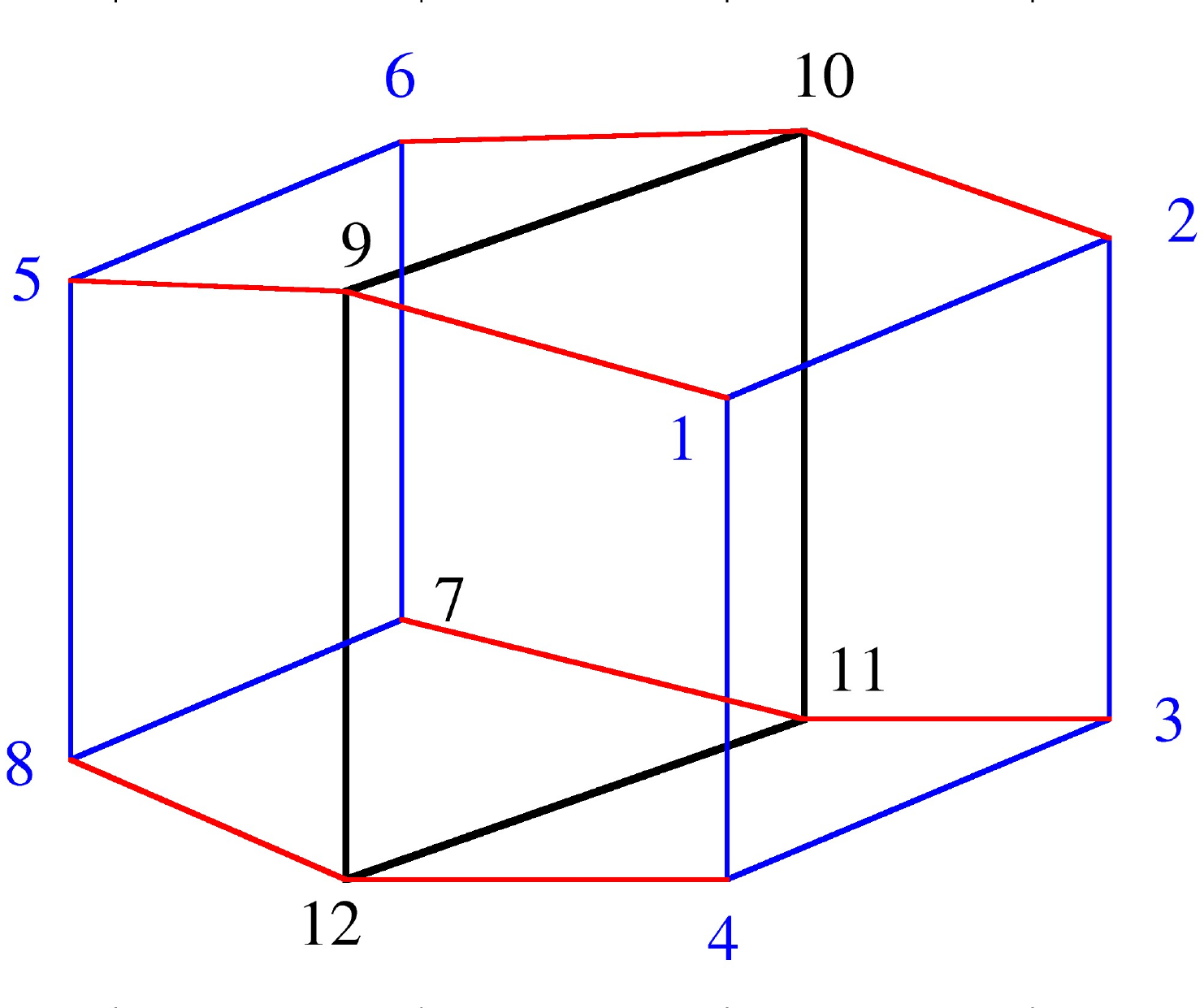}
	\caption{upper panel: Structure of $V_{12}$ molecule (CCDC - 1260088 after DFT geometric optimization), color code: grey - V, red - O, violet - As, white - H; lower panel: A schema depicting superexchange interactions between magnetic vanadium ions located in the corners of three squares: one black internal square (IS) - sites 9, 10, 11, 12 and two blue external squares (ES), sites 1, 2, 3, 4 and 5, 6, 7, 8.}
	\label{structure}
\end{figure}

There are few spherical dodecanuclear vanadium molecules \cite{Barra1992,Procissi2004,Barbour2006,Muller1991}. The most interesting are anions [V$_{12}$As$_8$O$_{40}$(HCO$_2$)]$^{n-}$ (with $n=3,5$) \cite{Muller1991}, which hereafter will be  called {\bf I} ($n=3$) and {\bf II} ($n=5$), as they feature two different valence states in the same structure (see Fig. \ref{structure}). {\bf I} contains six and {\bf II} eight unpaired electrons distributed over twelve vanadium ions. It is expected \cite{Gatteschi1993} that in the internal square (IS), which consists of sites 9, 10, 11 and 12 (Fig. \ref{structure}), there are only V$^{4+}$ ions (one unpaired 3d electron at each site), whereas in each of the external squares (ES), consisting of sites 1, 2, 3, 4 (ES1) and 5, 6, 7, 8 (ES2), there are one ({\bf I}) or two ({\bf II}) delocalized electrons (V$^{4+}/$V$^{5+}$ ions).  Low value of susceptibility at high temperature for {\bf II} \cite{Gatteschi1993} indicates that the electrons in the external squares in {\bf II} are strongly antiferromagnetically coupled. The magnetism of the molecules was investigated also with EPR measurements and modeled with a simplified Heisenberg model neglecting the delocalized nature of the electrons in the external squares \cite{Gatteschi1993}.  More advanced model has been developed later taking into account hoping of the electrons in the external squares \cite{Gatteschi1993a}. Yet, no parameters were fitted and only a general discussion has been carried out. 

To confirm a spin distribution for both molecules first the spin density is calculated by means of density functional theory (DFT). To this aim the X-ray determined structure is first geometrically optimized by DFT calculations, and then used do calculate spin density (More details on DFT calculations can be found in Appendix \ref{apdft}). The results are presented in Fig. \ref{spind}

\begin{figure}[h]
	\centering	
	\includegraphics[height=11cm]{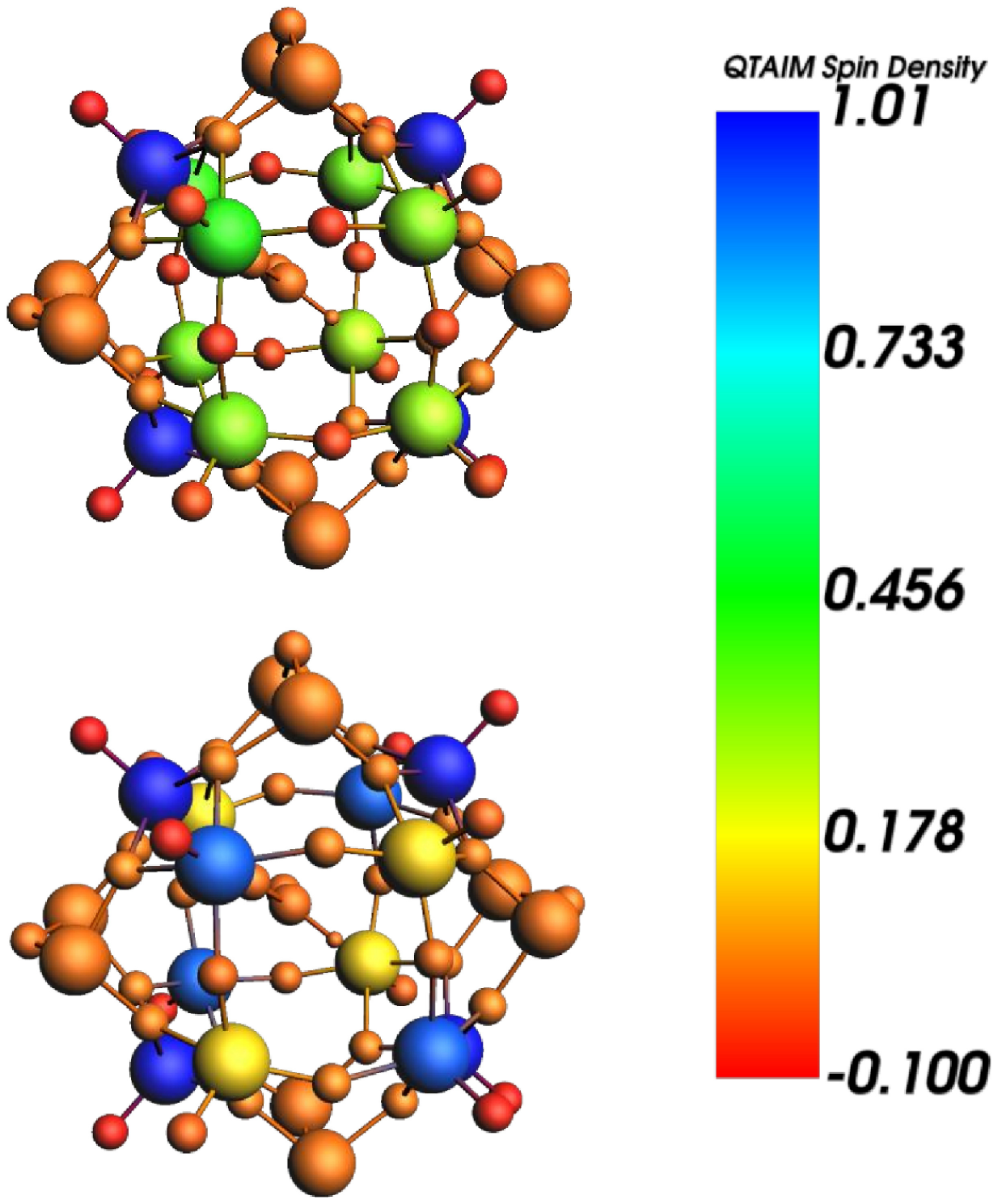}
	\caption{Spin density for molecule {\bf I} (upper panel) and  {\bf II} (lower panel).}
	\label{spind}
\end{figure}

As expected the spin density is everywhere close to zero (orange in Fig. \ref{spind}) except at vanadium sites. Only at some oxigens spin density is slightly negative indicating some leakage of the electrons into vanadium. 
For both molecules spin density at vanadium site in the IS is very close to 1 meaning that there is one unpaired electron at each site. In molecule {\bf I} the spin density at vanadium ions in the ES is almost the same and equal on average to $0.35$. It sums up to $1.4$ in each ES, which is more than expected $1$ - probably due to the leakage of electrons from neighboring oxygens which show small negative spin density.
In molecule {\bf II}  the spin density in the ES is more differentiated. At sites 1,3 and 6,8 the spin density is high ($0.92$ on average - blue color in Fig. \ref{spind}) and at the remaining sites 2,4 and 5,7 it is low ($0.135$ on average - yellow color in Fig. \ref{spind}). Spin density at vanadium ions in each of the ES sums up to $2.1$ which is very close to the expected value of $2$.

Thus, one can conclude that in both molecules IS contains vanadium ions with localized electrons corresponding to V$^{4+}$ oxidation state. In molecule {\bf I} each of the ES contains approximately one unpaired electron delocalized over four vanadium ions, whereas in molecule {\bf II} there are approximately two unpaired electrons in each of the ES localized mostly at sites 1,3 and 6,8. This picture is generally consistent with the original conjecture \cite{Gatteschi1993} except for molecule {\bf II} where a rather more uniform distribution was expected in the ES. The uneven distribution of itinerant electrons in the ES of molecule {\bf II} can be ascribed at this stage to Coulomb repulsion as the distance between sites 1 and 3 is larger than between sites 2 and 4 making the first pair of sites a preferred localization. Electron localization at sites 6 and 8 minimizes energy of Coulomb interaction with the itinerant electrons from the other ES and within the second external square as the distance between sites 6 and 8 is larger than between sites 5 and 7.

To model properly both molecules one needs to construct a Hamiltonian that will describe properly both itinerant and localized electrons. To this aim the most suitable seems to be t-J model in its full form corresponding to the expansion of the Hubbard model in the limit of strong on-site Coulomb repulsion \cite{Jedrak2011}. 

\begin{widetext}
\begin{eqnarray}
	{\cal H}&=&P\left\{\sum_{\langle i,j\rangle\in IS}J_1{\bf S}_i\cdot {\bf S}_j + \sum_{i=1}^4 J_2 \left({\bf S}_i + {\bf S}_{i+4}\right)\cdot {\bf S}_{i+8} +  g\mu_B\sum_{i=1}^{12}{\bf B}\cdot {\bf S}_i +
	\sum_{\langle i,j\rangle '\in ES, \sigma=\pm 1/2}t c_{i\sigma}^\dagger c_{j\sigma}\nonumber\right.
	\\
	&+&
	\sum_{(i,j)\in ES}\frac{n_i n_j q_e^2}{4\pi \epsilon_0 r_{ij}}+
	\sum_{\langle i,j\rangle\in ES}J_3 {\bf S}_i\cdot {\bf S}_j 
	+ \sum_{i,j,k\in ES, \sigma=\pm 1/2}\epsilon_{ijk}\left( c_{j\sigma}^\dagger c_{k\bar{\sigma}}^\dagger c_{j\bar{\sigma}}c_{i\sigma}- c_{j\sigma}^\dagger c_{k\bar{\sigma}}^\dagger c_{j\sigma}c_{i\bar{\sigma}} \right)
	\Bigg.\Bigg\}P\label{ham}
\end{eqnarray}
\end{widetext}

Operator $P=\prod_i(1-n_{i\frac{1}{2}}n_{i-\frac{1}{2}})$, where $n_{i\sigma}=c_{i\sigma}^\dagger c_{i\sigma}$, eliminates states with double occupied sites which in the case of vanadium ions in square pyramid coordination with vanadyl bond are strongly disfavored \cite{Kaul2005}. The first term describes superexchange interactions in the IS. The symbol $\langle i,j\rangle$ stands for all nearest neighbor (nn) bonds, which are counted only once.  The second term describes superexchange interactions between electrons in the IS and itinerant electrons in the ES. It is assumed that $S=1/2$ spin operator ${\bf S}_i$ vanishes when acting on an empty site. The third term stands for Zeeman interaction with the magnetic field ${\bf B}$. The forth term describes hoping of the electrons in the ES. The symbol $\langle i,j\rangle '$ stands for all nn bonds, but counted twice, for instance both $(1,2)$ and $(2,1)$ pairs are present in the sum. The fifth term describes Coulomb interactions between itinerant electrons. $n_i=\sum_\sigma n_{i\sigma}$ is an occupation operator, $q_e$ is electron charge, $\epsilon_0$ is dielectric permeability of vacuum and $r_{ij}$ stands for distance between two itinerant electrons. Symbol $(i,j)$ stands for all the pairs (counted once) not only nn. The remaining two terms are present only for molecule {\bf II}. The sixth term describes superexchange interactions between itinerant electrons in the ES. The last term is a so called three site hoping term and describes correlated hoping of electrons with the opposite polarization.

The last term is often omitted, but it is of the same order as the exchange terms \cite{Jedrak2011} and it is demonstrated to be important in similar systems \cite{Suaud2015}. Besides, the magnetic measurements suggest strong antiferromagnetic coupling of electrons in the ES and DFT results indicate that these electrons are mostly apart from each other. Such situation cannot take place with only superexchange interaction as superexchange requires proximity of the electrons. 
The correlated hoping described by the last term of Hamiltonian (\ref{ham}) can be divided into two types. In the first type one nn spin distribution in the ES is changed into another nn distribution, whereas the second type mixes diagonal and nn distributions (See an example in Appendix \ref{apham}). Both forms enforce antiferromagnetic alignment of the itinerant spins, but the second one keeps electrons at a distance. Therefore, in what follows parameter $\epsilon_{ijk}$ will take two values: $\epsilon_1$ for hoping between nn distributions and $\epsilon_2$ for hoping between nn and diagonal distributions.

In similar systems the orbital energy is often added to the Hamiltonian \cite{Cardona-Serra2013,Cardona-Serra2015,Suaud2015}, however due to the symmetry of the molecule and a constant number of itinerant electrons such a term would only add some constant to the Hamiltonian and therefore is omitted.

The localisation of the unpaired electrons in the IS predicted by DFT calculations can be ascribed to the lower orbital energy in the IS. It is assumed at this stage that even higher temperature cannot induce electron transfer from the IS to the one of the ES. The transfer of the electrons between the ES through the occupied IS is forbidden due to the large energy gap between single and double occupied vanadium sites \cite{Kaul2005}. Therefore for vanishing electric field only distributions 1-4-1 (molecule {\bf I}) and 2-4-2 (molecule {\bf II}) are considered, which stand for 4 unpaired electrons in the IS and 1 (for {\bf I}) or 2 (for {\bf II}) unpaired electrons in each of the ES.

\section{Fitting the experimental data \label{fitting}}
To obtain a realistic values of parameters in Hamiltonian
(\ref{ham}) one should fit the available experimental data \cite{Gatteschi1993}. However a number of parameters makes this task difficult. Therefore first molecule {\bf I} will be fitted, as it requires only 4 parameters. Then the parameters obtained for {\bf I} will be used in fitting {\bf II}. If the fitting gives more than one unique result the optimal set of parameters will be chosen by comparing spin density calculated by DFT with the electron distribution obtained for a given set of parameters with Hamiltonian (\ref{ham}). Such a procedure has  already been proven to be very efficient \cite{Kozlowski2017,Notario2018}.

\begin{figure}[h]
	\centering	
	\includegraphics[height=6cm]{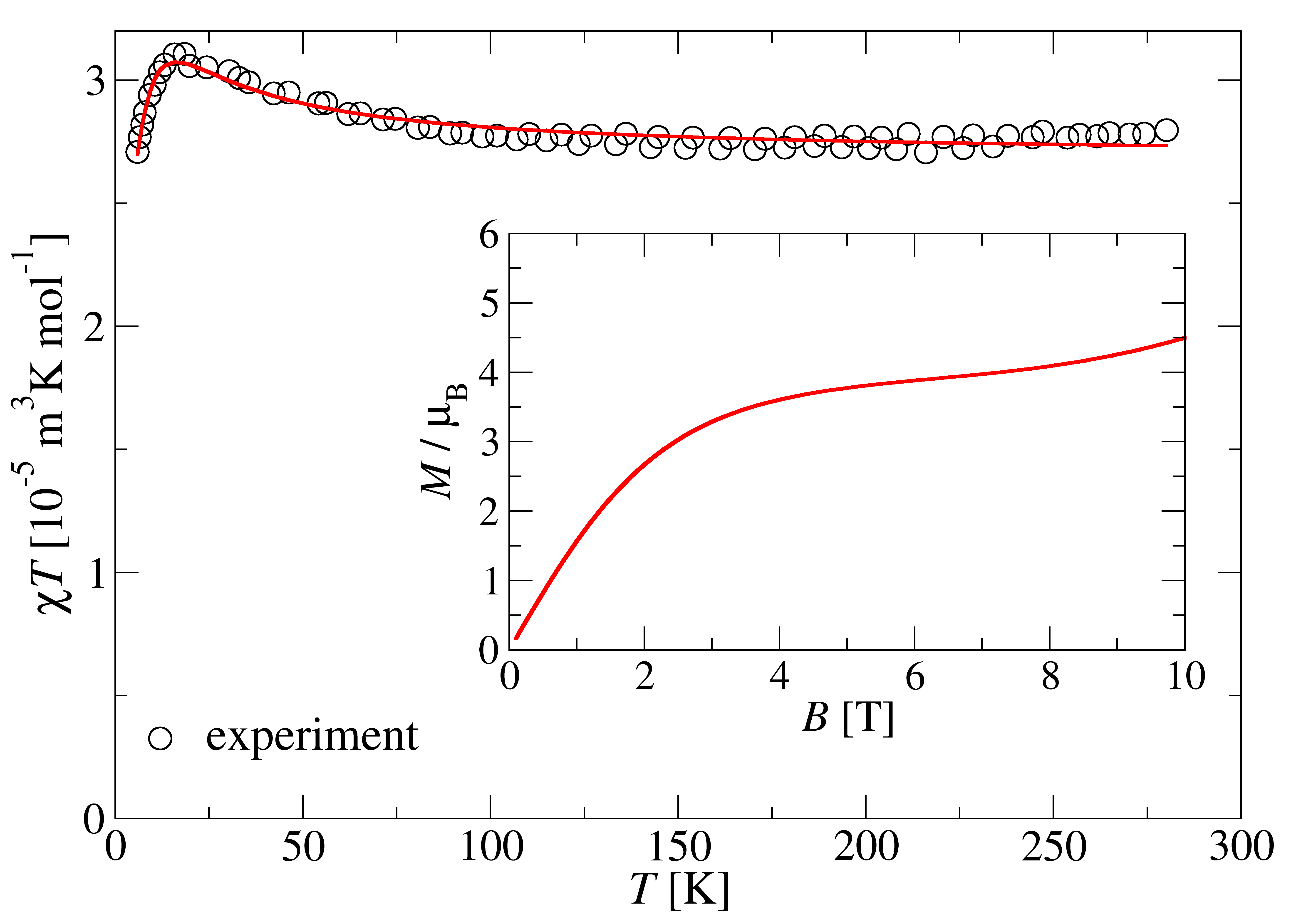}	
	\includegraphics[height=6cm]{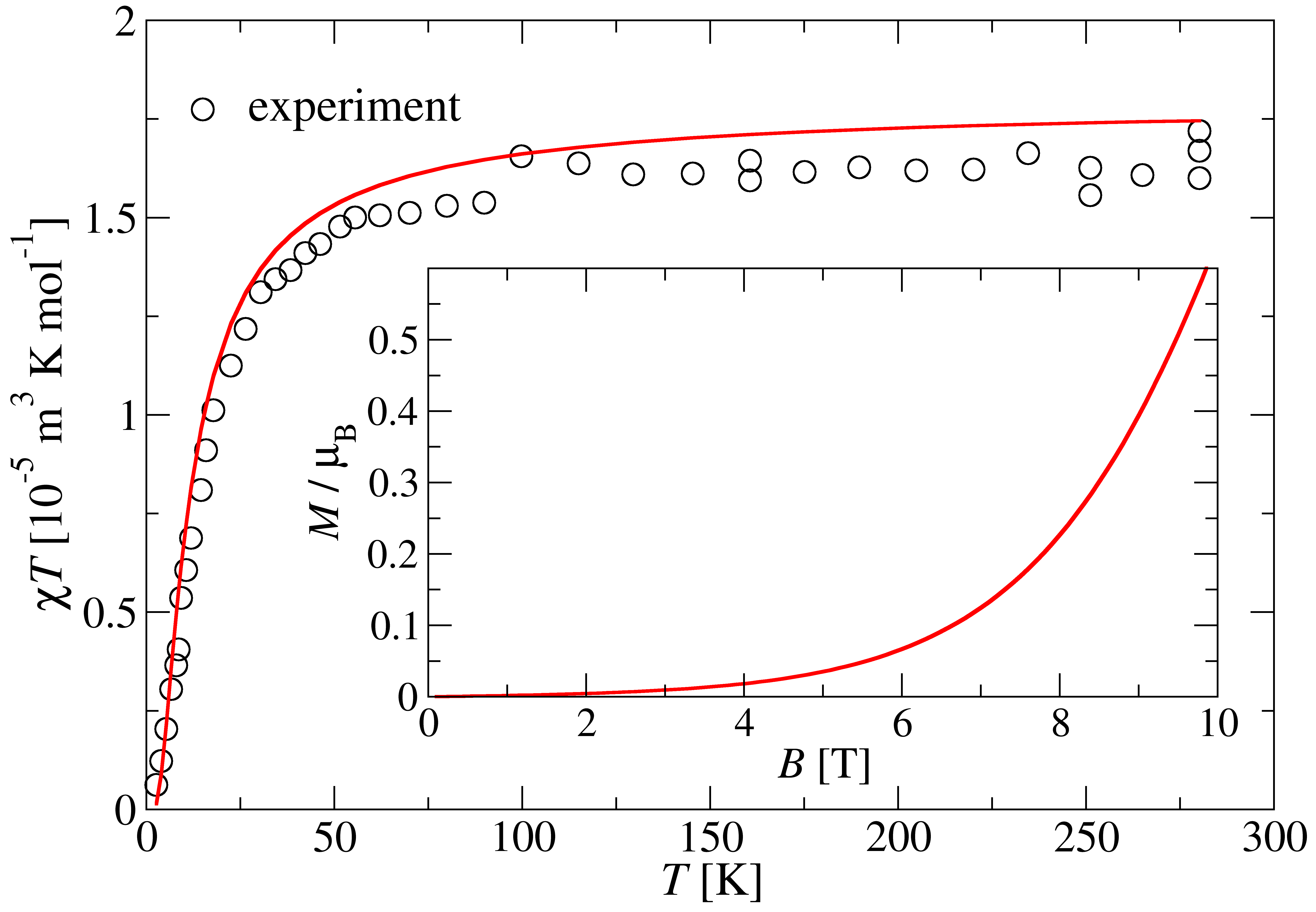}
	\caption{Temperature dependence of $\chi T$ ($B=0.1$) for {\bf I} (upper panel) and {\bf II} (lower panel) \cite{Gatteschi1993} with the optimal fits (red lines) and the theoretical prediction of magnetic field dependence of magnetization at $T=2$ K (in the inserts).}
	\label{fit}
\end{figure}

One of the drawbacks of the modeling presented in paper \cite{Gatteschi1993} is the fact that the value of $g$ from EPR does not agree with that obtained from the fitting of magnetic susceptibility. It can be demonstrated that the use of more advanced Hamiltonian (\ref{ham}) does not help, and the difference between the fitted and EPR value is even larger. 
This problem can be solved by correcting the magnetic data from paper \cite{Gatteschi1993} for diamagnetism. After extracting the diamagnetic contribution of the V$_{12}$ molecule, the counter-ions and the solvent \cite{Muller1991} from the experimental data the fitted and the EPR values agree perfectly. Thus, in what follows the g value is fixed to the EPR value of $1.95$ and the experimental data presented in all figures are corrected for diamagnetism. Thus,  for {\bf I} only three parameters have to be determined. Fig. \ref{fit} presents the best fit of magnetic molar susceptibility $\chi$ obtained for $J_1/k_B=14.53$ K, $J_2/k_B=-57.86$ K and $|t/k_B|\geq 2000$ K.

The value of $|t/k_B|$ above $2000$ K does not influence the shape of the susceptibility curve (see Fig. \ref{tdep} in Appendix \ref{apham}) Due to the square topology the negative and positive values of $t$ generate the same energy spectrum of Hamiltonian (\ref{ham}) \cite{Suaud2015}. The ground state of ${\bf I}$ is a singlet $S=0$, but two lowest excited states are very close to the ground state: $S=1$ ($0.04$ K above the ground state) and $S=2$ ($0.29$ K above the ground state). Thus, the low temperature behavior of molecule {\bf I} is determined by quasi degenerated states $S=0$, $S=1$, and $S=2$.

For fitting molecule {\bf II} parameters $J_1$ and $J_2$ were fixed to the values obtained for {\bf I} and $t$ is limited to values $|t/k_B|\geq 2000$ K. To obtain a good fit to susceptibility it is enough to use only exchange term with coupling $J_3/k_B\approx 5000$ K. However this term strongly favors nn configurations of itinerant electrons which makes it impossible to obtain the electron distribution concordant with the DFT results. A similar situation appears when a three site correlated hoping of type one is introduced. Therefore in the fitting procedure it is assumed that $\epsilon_1=0$ and only three parameters $t$, $J_3$ and $\epsilon_2$ are fitted. The optimal values, which give simultaneously a good fit to susceptibility and the electron distribution concordant with the DFT results are the following: $t/k_B=\pm 2000$ K and $\epsilon_2/k_B=\mp 2000$ K (see Fig. \ref{fit}). Their signs are correlated, that is if $t$ is positive $\epsilon_2$ is negative and vice versa. The value of $J_3$ cannot be determined since due to the preferred diagonal distribution of the itinerant electrons any sufficiently small value of $J_3$ gives the same fit to susceptibility and the same electron distribution. Larger values of $J_3$ lead to wrong electron distribution.
The ground state of molecule {\bf II} is an $S=0$ singlet and the first excited state $S=1$ lays $14.5$ K above the ground state.

The new fits have many advantages over the fits obtained with the simplified models \cite{Gatteschi1993}: they are better (closer to experimental values), explain dynamics, agree with the EPR measurements and the DFT calculations. In table \ref{locdftham} one can find comparison between spin densities obtained by DFT and probability of finding a spin at a given site obtained from Hamiltonian \ref{ham} with optimal parameters. For this comparison the spin densities of vanadium ions are rescaled so that their sum over the both ES gives value $2$ for {\bf I} and $4$ for {\bf II}.

\begin{table}[h]
	\small
	\caption{\ Rescaled spin densities obtained by DFT and probabilities to find an electron at a given site obtained with Hamiltonian (\ref{ham}) for {\bf I} and {\bf II} at $T=2$ K and $B=0$}
	\label{locdftham}
	\begin{tabular*}{0.5\textwidth}{@{\extracolsep{\fill}}ccccc}
		\hline
		number of V site & DFT ({\bf I})& Ham ({\bf I})& DFT ({\bf II}) & Ham({\bf II}) \\
		\hline
		1 & 0.31 &0.27& 0.87&  0.89\\
		2 & 0.23 & 0.24& 0.12 & 0.11\\
		3 & 0.22 & 0.25&0.88 & 0.89\\
		4 & 0.24 & 0.25&0.13&  0.11\\
		5 & 0.30 & 0.22&0.14&  0.11\\
		6 & 0.25 &0.26& 0.87 & 0.89\\
		7 & 0.20 &0.25& 0.13 & 0.11\\
		8 & 0.25 &0.26& 0.86&  0.89\\
		\hline
	\end{tabular*}
\end{table}

\section{Impact of magnetic field}
\begin{figure}[h]
	\centering	
	\includegraphics[height=6cm]{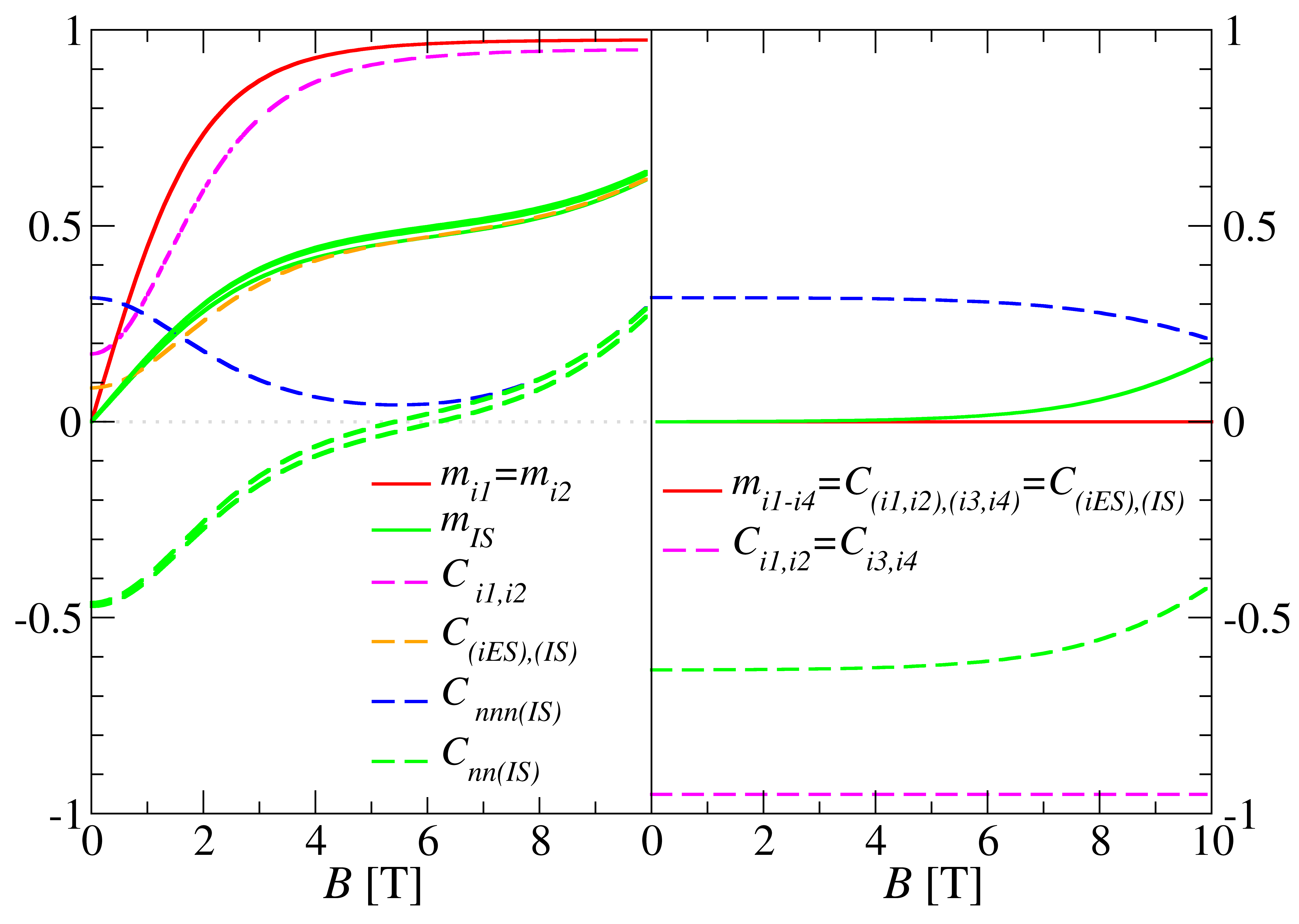}
	\caption{Local magnetizations and corelations at $T=2$ K for molecule {\bf I} (left) and {\bf II} (right).}
	\label{coref0}
\end{figure}

The localizations of the electrons presented in table \ref{locdftham} do not change if magnetic field (up to 10 T) is applied. However local magnetizations and correlations do change especially for molecule {\bf I}. The results are presented in Fig. \ref{coref0}. Local magnetizations and correlations are defined as follows:
\begin{equation}
	m_i=g\langle S_i\rangle,~~C_{i,j}=g^2\langle S_iS_j\rangle ~~i=1,...,12~~{\rm or}~~i=i1,...,i4
\end{equation}
where $\langle ...\rangle$ stands for thermal average and index $i1,...,i4$ points to the itinerant electrons and not to particular sites of the molecule.

Local magnetizations in the IS (solid green lines in Fig. \ref{coref0}) increase with magnetic field in both molecules, but much slower in molecule {\bf II}. In the IS at $B=0$  correlations between nn (green broken lines in Fig. \ref{coref0}) are negative and those between next nearest neighbors (nnn) (blue broken lines in Fig. \ref{coref0}) are positive, which agrees with the antiferromagnetic coupling $J_1$ in the IS. However in molecule {\bf I} all the electrons in the IS are positively correlated for magnetic fields above $6$ T. In {\bf II} even in high magnetic fields correlations in the IS conserve their character, only their absolute value is a bit smaller.
The biggest difference between the molecules concerns the behavior of the itinerant electrons. In molecule {\bf I} itinerant electrons have small positive correlation at magnetic field $B=0$ which quickly increases for higher fields (magenta broken line in Fig. \ref{coref0}). In molecule {\bf II} itinerant electrons within the ES are strongly antiferromagnetically coupled, whereas there is completely no correlation between itinerant electrons  from different ES. Due to this strong coupling the itinerant electrons in {\bf II} are not magnetized even in high magnetic field, whereas in {\bf I} they quickly respond to magnetic field giving rise to high values of magnetisation (red lines in Fig. \ref{coref0}.) The correlation in {\bf I} between itinerant electrons in the ES and the electrons in the IS is the same independently of the distance between the electrons and increases with the magnetic field (orange broken line in Fig. \ref{coref0}). In molecule {\bf II} correlations between the ES and the IS vanish for any value of the magnetic field.

\section{Impact of electric field}\label{elfield}
To study the influence of uniform electric field $\bf E$ the following Hamiltonian has to be added to Hamiltonian (\ref{ham}):

\begin{equation}
	{\cal H}_{ef}=q_e\sum_{i\in ES}{\bf r}_i\cdot{\bf E}\label{hamef}
\end{equation}
where ${\bf r}_i$ is a position vector of vanadium sites in the ES with unpaired electrons.

If the electric field is applied parallel to the ES, e.g. along sites 1 and 2 or 4 and 2 the change of orbital energy induced at neighboring vanadium sites belonging to the IS and the ES is the same. Thus, there should be no electron transfer between the IS and the ES and only distributions 1-4-1 for {\bf I} and 2-4-2 for {\bf II} have to be considered. However, the application of the electric field perpendicular to the ES may lead to the situation in which the orbital energy at vanadium in the IS is higher than at the vanadium in one of the ES causing the electron transfer from the IS to the ES, and then from another ES to the IS to fill the electron gap, thus leading to the effective electron transfer between the ES. In such a situation the following distributions should be considered: 1-4-1, 2-4-0, 0-4-2 for molecule {\bf I} and 2-4-2, 1-4-3, 0-4-4, 3-4-1, 4-4-0 for molecule {\bf II}. There is no need to consider distributions in which in the IS there is less than four unpaired electrons as all the DFT calculations in the considered electric field range give all the vanadium spin densities in the IS equal to 1.

\subsection{Electric field parallel to the ES}
\subsubsection{Molecule {\bf I}}
Due to the symmetry of the molecule there are generally two nonequivalent directions of the electric field applied parallel to the ES: along sites 4 and 2 and along sites 1 and 2. In this section only the direction 4-2 will be considered in detail, as for this direction the impact of electric field on magnetism of molecule {\bf I} measured by variation of the $\chi T$ dependence on temperature and magnetisation dependence on magnetic field is the largest (compare Fig. \ref{chimag42} with Fig. \ref{chimagn12} in Appendix \ref{apham}).

As can be seen in Fig. \ref{chimag42} both magnetic susceptibility and magnetization are changed by electric field. The change is gradual up to $E=10$ V/nm, though very small for fields between $5$ and $10$ V/nm. For $E>10$ V/nm no more change can be observed. The effect is detectable below $T=100$ K. 

\begin{figure}[h]
	\centering	
	\includegraphics[height=6cm]{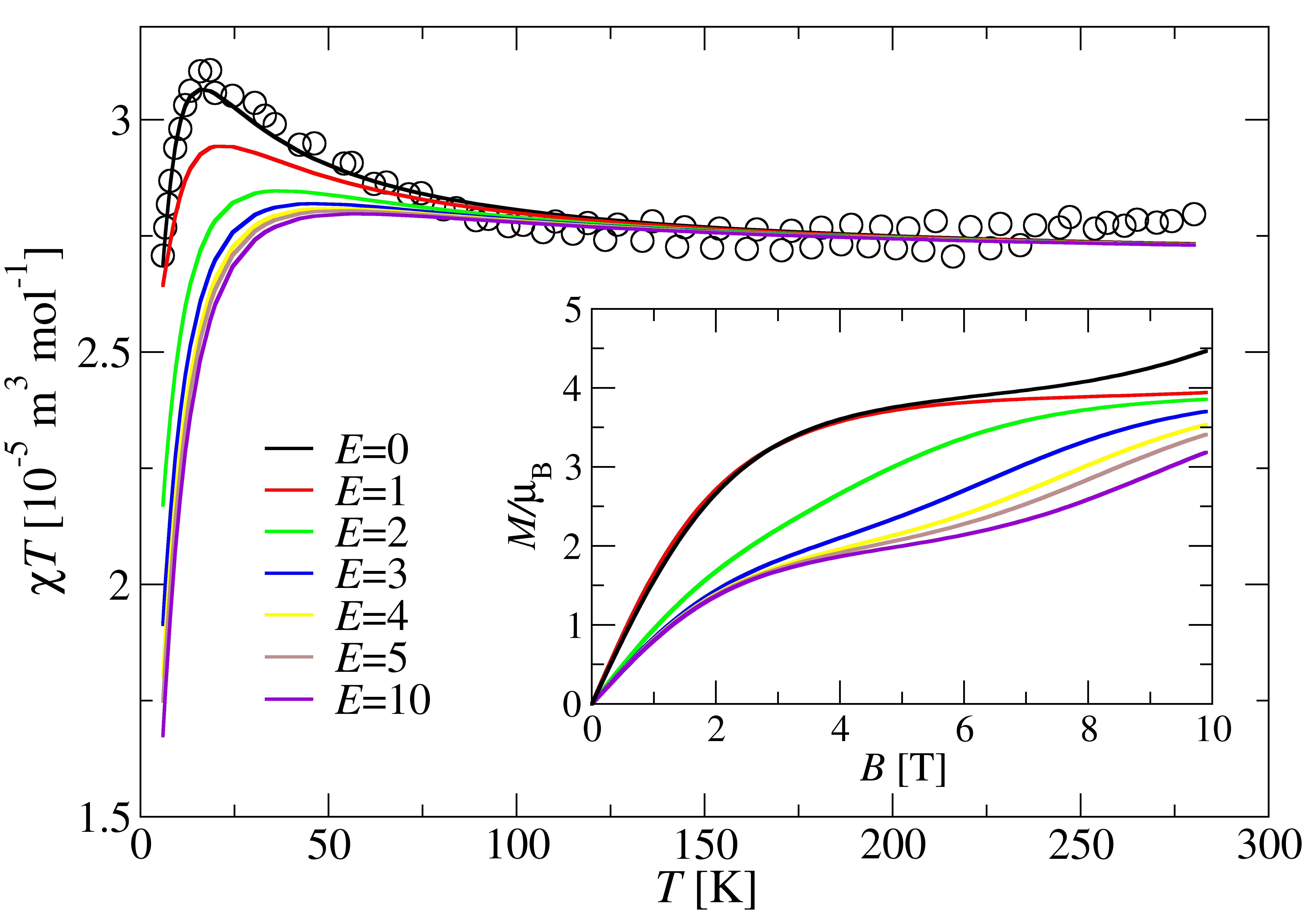}
	\caption{Temperature dependence of $\chi T$ ($B=0.1$ T) and magnetic field dependence of magnetization $M$ for molecule {\bf I} in different electric fields $E$ [V/nm] applied along sites 4 and 2. Empty circles stand for experimental results \cite{Gatteschi1993} at field $E=0$.}
	\label{chimag42}
\end{figure}
\begin{figure}[h]
	\centering	
	\includegraphics[height=6.5cm]{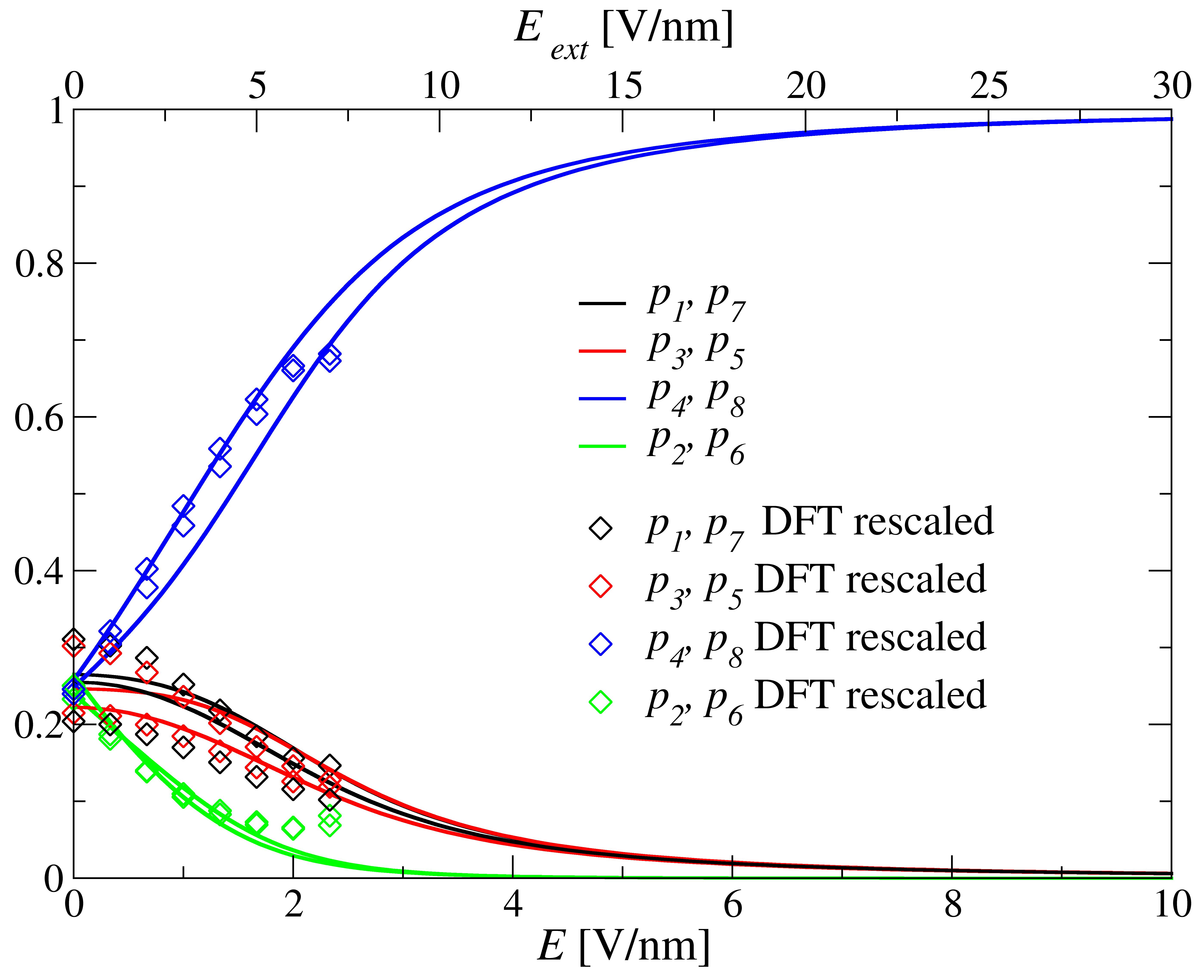}
	\caption{Electric field dependence of probability $p_i$  to find an unpaired electron at a given vanadium site at $T=2$ K and $B=0$ for molecule {\bf I}. Diamonds stand for rescaled spin densities obtained by DFT in a given external electric field $E_{ext}$ (upper scale). Electric field is applied along sites 4 and 2.}
	\label{loc42-goc-B0}
\end{figure}

In Fig. \ref{loc42-goc-B0} the localisation of the electrons in the ES is shown.
It can be seen that due to the influence of electric field the itinerant electrons from the ES move to sites 4 and 8. Thus, they are forced to interact mostly with the localized electron at site 12 (see lower panel in Fig. \ref{structure}).  
For the electric field applied along sites 1 and 2 the itinerant electrons move to sites 1, 4 and 5, 8, that is they become less localized than with the field applied along sites 4 and 2 (see Fig. \ref{loc12I} in Appendix \ref{apham}).

The electric field $E$ in Hamiltonian (\ref{hamef}) is a field experienced at a given vanadium site and not the applied external field. To estimate the screening effect spin density was estimated by means of DFT in various external electric fields. The results can be seen in Fig. \ref{loc42-goc-B0}. To obtain a good agreement between DFT and Hamiltonian calculations the external electric field used in DFT should be divided by $3$, which suggests that due to the screening effect the local field is three times smaller than the applied external electric field. The DFT results are shown only for external fields below $7$ V/nm since for higher fields some of the spin density is moved out of the vanadium cores changing the electronic structure to the form that cannot be modeled by Hamiltonian (\ref{ham}). Below $E_{ext}=7$ V/nm according to DFT calculations the molecule remains fully intact and only electron redistribution among vanadium cores can be observed.

\begin{figure}[h]
	\centering	
	\includegraphics[height=6cm]{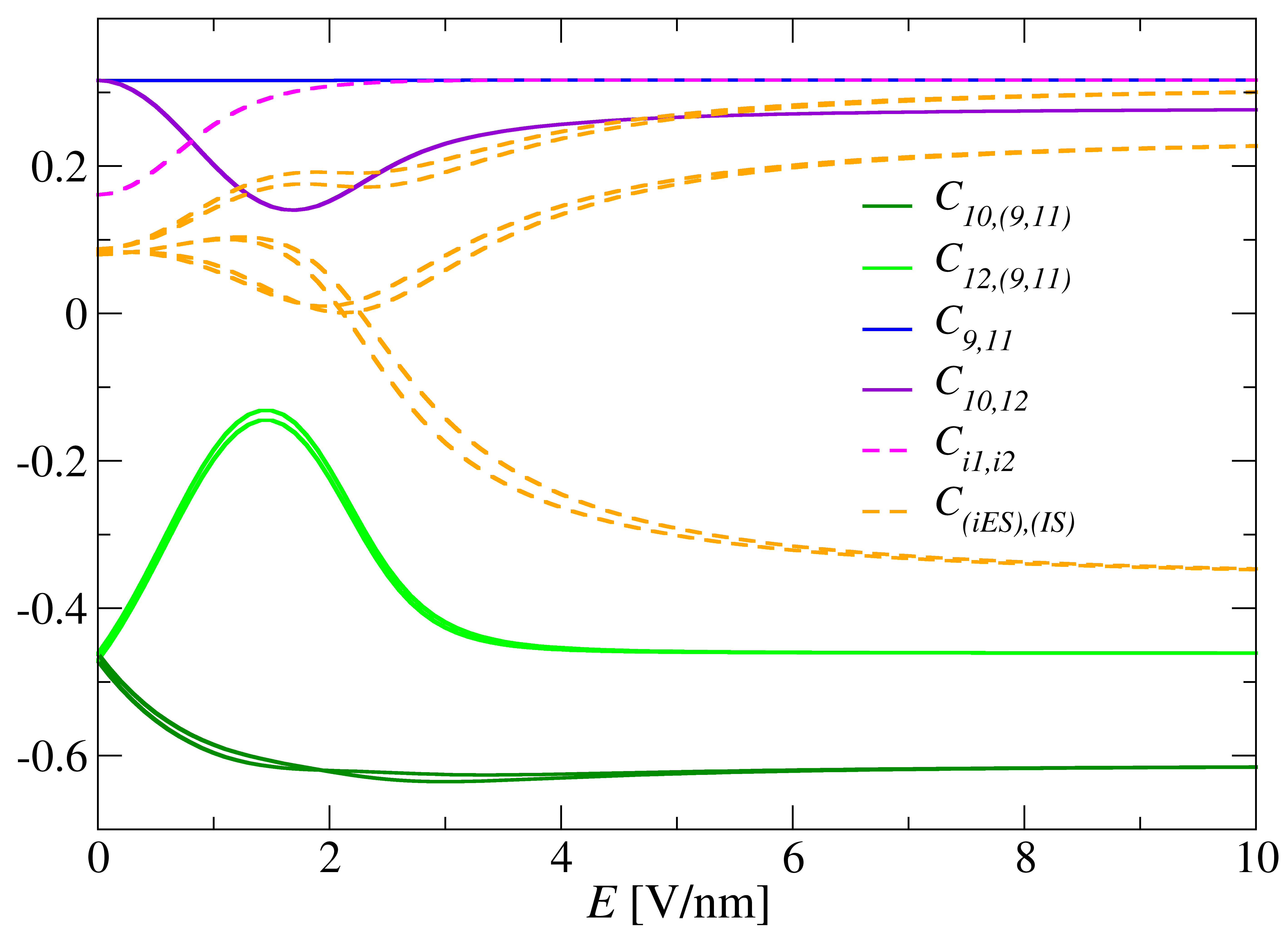}
	\caption{Magnetic correlations for molecule {\bf I} versus electric field applied along sites 4 and 2 at $T=2$ K and $B=0$.}
	\label{corIgoB0}
\end{figure}

The relocation of the itinerant electrons influences their interactions with other electrons which is reflected in the correlations (see Fig. \ref{corIgoB0}).
It can be seen that correlations between electrons in the IS engaging site $12$ show an extremum for the field slightly smaller than $2$ V/nm. The correlation between the itinerant electrons in the ES and the electrons in the IS split into three branches. The highest one corresponds to the correlation with the nn, which is usually site $12$ as the electric field increases. The second positive branch corresponds to the interaction with the most distant site (for large field it is site $10$). The negative branch corresponds to correlations with nnn. The correlation between itinerant electrons increases, but not much. All this behavior can be ascribed to the increasing localisation of the itinerant electrons as the electric field rises.

\begin{figure}[h]
	\centering	
	\includegraphics[height=6cm]{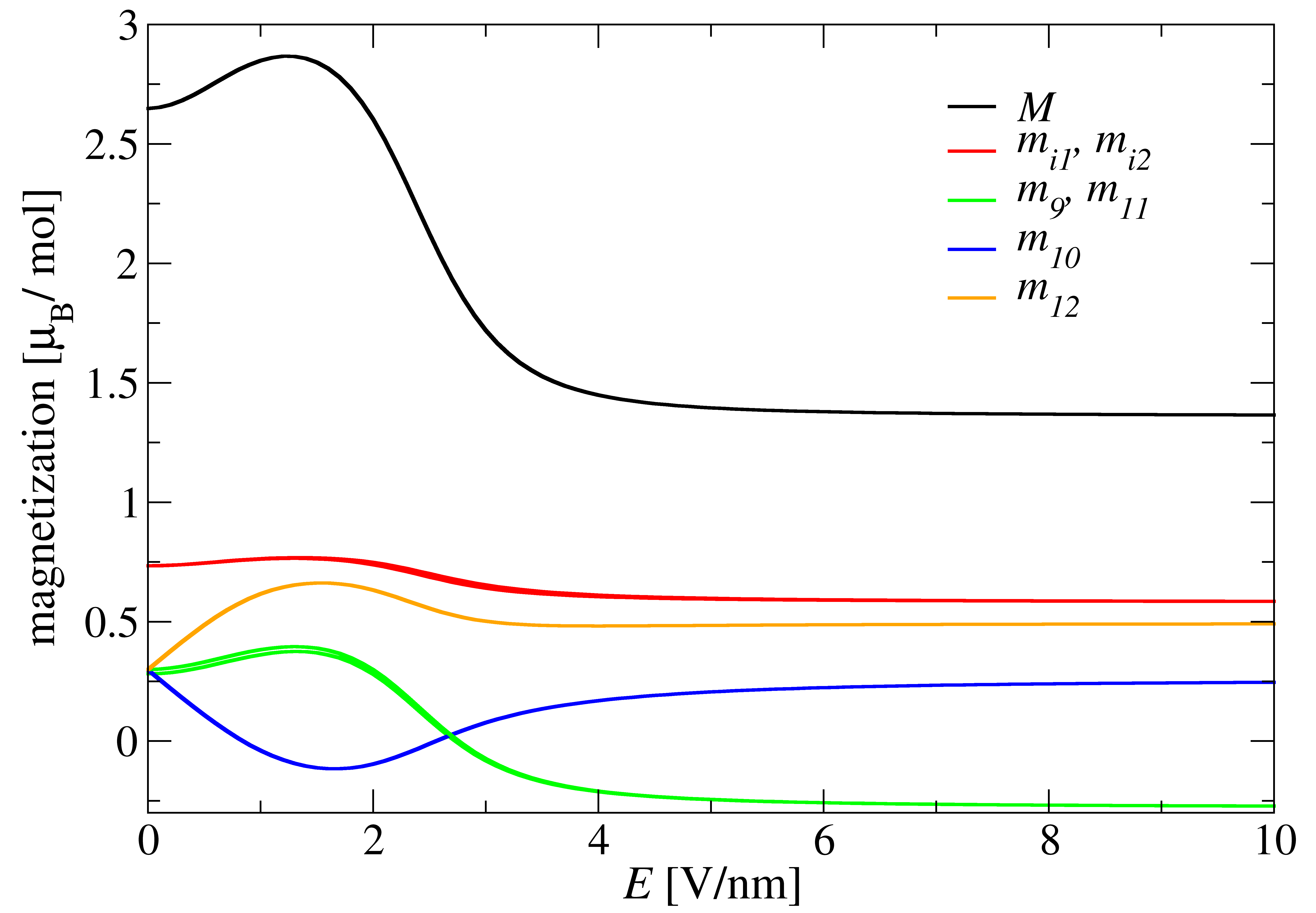}
	\caption{Global ($M$) and local ($m_i$) magnetisations for molecule {\bf I} at $T=2$ K and $B=2$ T. Electric field is applied along sites 4 and 2.}
	\label{magIgoct2b2}
\end{figure}

To see the influence of the electric field on local magnetizations one needs to apply some non-zero magnetic field (see Fig. \ref{magIgoct2b2}).
All local magnetisations $m_i$ and the global magnetisation $M=\sum_i m_i$ exhibit an extremum for the electric field between $1$ and $2$ V/nm. Besides, local magnetisations in the IS split into three branches, very much the same like the correlations between itinerant electrons and the electrons in the internal square (orange curves in Fig. \ref{corIgoB0}). The electric field dependence of the global magnetisation marks a spin crossover at around $2$ V/nm. As the electric field increases the spin state changes from high to low. The electric field dependence of correlations at magnetic field $B=2$ T is qualitatively the same like that presented in Fig. \ref{corIgoB0} for $B=0$, though some of the extrema are better pronounced.

\begin{figure}[h]
	\centering	
	\includegraphics[height=6cm]{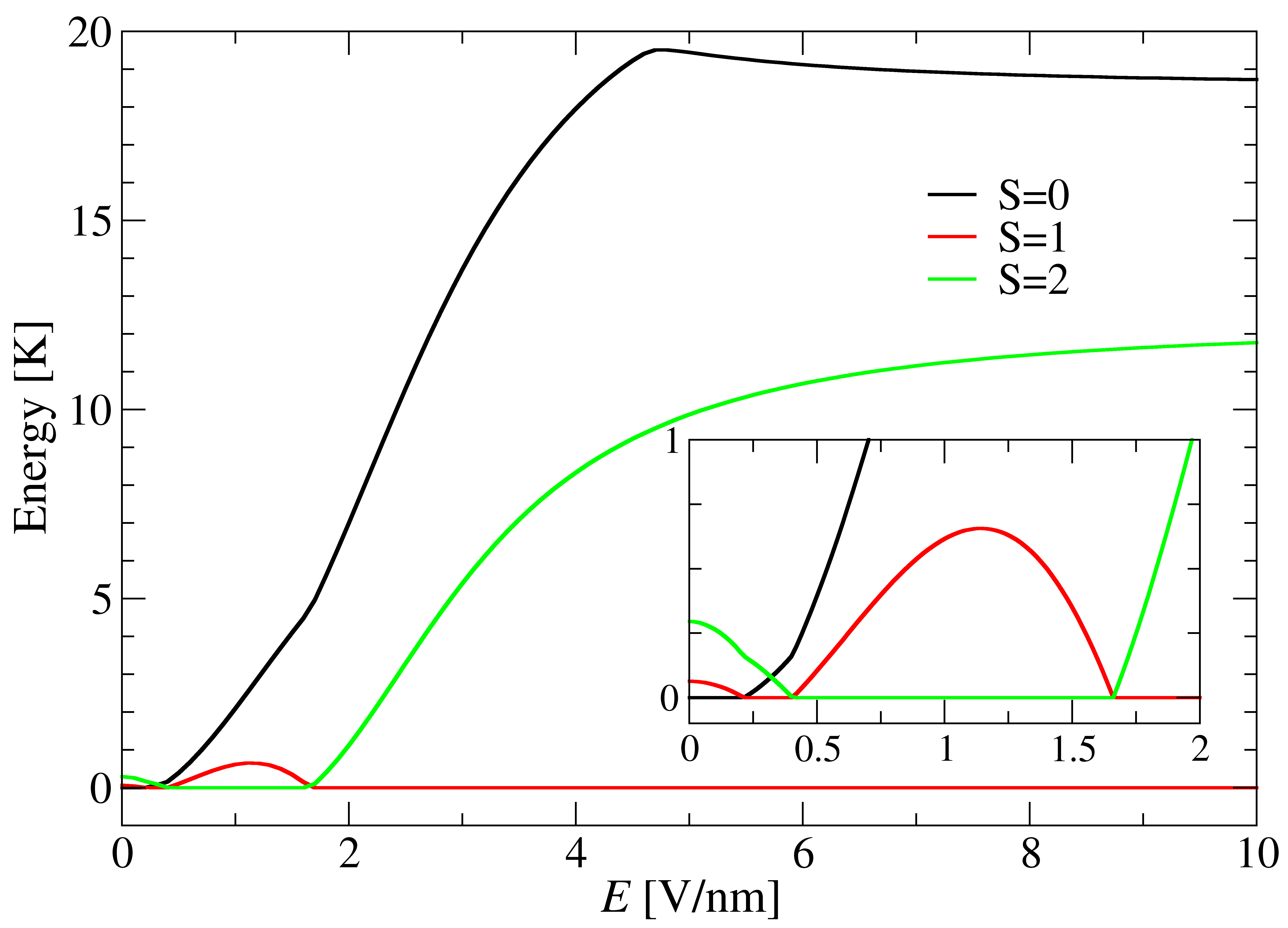}
	\caption{The ground state and two lowest excited states of molecule {\bf I} as a function of electric field $E$ along sites $4$ and $2$ at $B$=0. The energy is rescaled so that the ground state has energy equal to $0$.}
	\label{groundsI}
\end{figure}

The change of low temperature magnetic behavior depicted in the last two figures is caused by the change of the ground state which is presented in Fig. \ref{groundsI}.
Three energy levels $S=0$, $S=1$, $S=2$ are quasi degenerated at $E=0$ and split with increasing electric field. For higher fields their separation of $11$ K between $S=1$ and $S=2$ and $7$ K between $S=2$ and $S=0$ seems to saturate. The ground state changes from $S=0$ through $S=2$ to $S=1$. Thus, one can observe a switching of the ground state of the molecule by applied electric field leading to a perceptible change of magnetic behavior.

\subsubsection{Molecule {\bf II}}
\begin{figure}[h]
	\centering	
	\includegraphics[height=6.5cm]{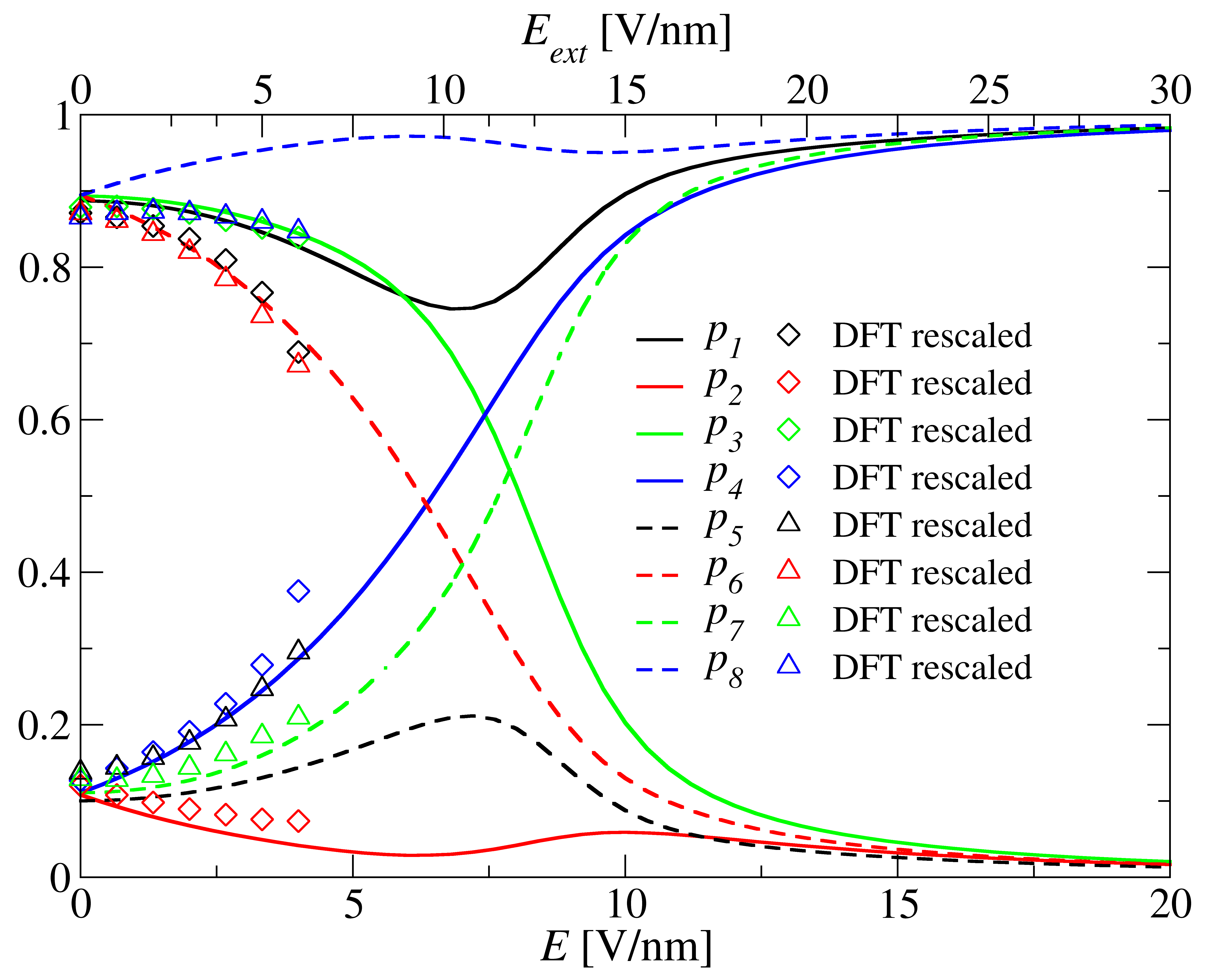}
	\caption{Electric field dependence of probability $p_i$  to find an unpaired electron at a given vanadium site in the ES of molecule {\bf II} at $T=2$ K and $B$=0. Solid and broken lines mark quantities corresponding to the different ES. Symbols stand for rescaled spin densities obtained by DFT in external electric field $E_{ext}$ (upper scale). Electric field is applied along sites $4$ and $2$.}
	\label{loc42-B0II}
\end{figure}

Molecule {\bf II} should have similar anisotropy with respect to the electric field as molecule {\bf I}. Thus, one could expect the strongest magneto-electric effect for the field applied along sites $4$ and $2$. However, contrary to {\bf I} no influence of the electric field (up to $20$ V/nm) on magnetic susceptibility and magnetisation has been found. The same concerns other directions of the field parallel to the ES. Yet, one can demonstrate that the itinerant electrons in the ES change their positions under influence of electric field (see Fig. \ref{loc42-B0II}).

If the electric field is applied along sites $4$ and $2$ the itinerant electrons move to sites $4$ and $7$ leaving almost empty sites $3$ and $6$. Thus the preferred localisation changed from sites $1$, $3$ and $6$, $8$ to sites $1$, $4$ and $7$, $8$.

The best, though not perfect fit, to the DFT results is obtained with the rescaling factor $1.5$ between external $E_{ext}$ and local electric field $E$. The largest differences can be found for sites 1, 5 and 8. However, the overall character of these curves cannot be fully determined from DFT as for fields larger than $E_{ext}=6$ V/nm the DFT results indicate that the spin density is driven out of vanadium cores and thus the system cannot be modeled by Hamiltonian (\ref{ham}). For external fields smaller than $6$ V/nm the molecule remains intact and only the redistribution of electrons among vanadium cores is observed.
The value of the rescaling factor different than for molecule {\bf I} may seem strange, but it may be due to the fact, that the screening effect may be mostly due to the itinerant electrons, which though more numerous in molecule {\bf II} are less mobile due to strong repulsion and exchange transfer ($\epsilon_2$), which keep them at a distance.
For the direction of the electric field along sites $1$ and $2$ the preferred localization changes to sites $1$, $4$ and $5$, $8$ (see Fig. \ref{loc12II} in Appendix \ref{apham}).
Nevertheless the shifting of the electrons has also almost no influence on correlations. This is probably the effect of a very strong antiferromagnetic coupling of the itinerant electrons induced by a large value of parameter $\epsilon_2$.

\subsection{Electric field perpendicular to the ES}
Since the application of electric field perpendicular to the ES can lead to different orbital energies in different ES and thus to the effective electron transfer between the ES  the calculations with Hamiltonian (\ref{ham}) extended by Hamiltonian (\ref{hamef}) are considered in the larger space corresponding to different electron distributions (see the beginning of section \ref{elfield}). Both methods DFT and Hamiltonian calculations provide evidence for an effective electron transfer between the ES. However the critical fields at which these transitions happen are different for both methods. In order to keep the relation between the external electric field as implemented in DFT and internal (local) electric field in the Hamiltonian the same like for the parallel application of the electric field the corrections to orbital energy in the new, induced by electric field, electron distributions are introduced. Such an approach can be justified by the fact that the Coulomb interactions between nn itinerant electrons are at least partially accounted for by superexchange interactions. Because the number of nn superexchange interactions between the itinerant electrons increases as the electrons are transferred from one ES to the other more of the Coulomb interactions are accounted for by superexchange and therefore the orbital energy should be modified to compensate this change. To this aim one (for molecule {\bf I}) and two (for molecule {\bf II}) parameters are added to account for this effect. The values of these parameters are chosen in such a way that the relation between external and local electric fields as found in previous subsection is conserved.

\subsubsection{Molecule I}
The application of the electric field along sites 1 and 5 induces the change of the 1-4-1 distribution into 2-4-0 (or 0-4-2). The parameters obtained from the fitting at zero electric field are kept intact. Two new parameters appear: $\epsilon_2$, due to exchange transfer in one of the ES with two electrons and $\Delta E_o$ which accounts for the change of the orbital energy in new electron distribution. Both these parameters appear only in 2-4-0 (0-4-2) distribution.  $\Delta E_o$ is just added to the Hamiltonian (if distributions 2-4-0 or 0-4-2 are active) and $\epsilon_2$ is defined like for molecule II. To obtain a good agreement between spin densities obtained by DFT and probability of occupation of vanadium sites calculated with the Hamiltonian and to keep the relation between the external and local electric field the same like for parallel orientation of the electric field the new parameters should take the values: $\epsilon_2=-1000$ K, $\Delta E_o=-7200$ K. In distributions 2-4-0 and 0-4-2 in principle also the superexchange interaction between the itinerant electrons can take place. However the DFT shows that itinerant electrons are kept at a distance. Thus the superexchange interaction is ineffective and therefore it is assumed that $J_3=0$.

\begin{figure}[h]
	\centering	
	\includegraphics[height=7cm]{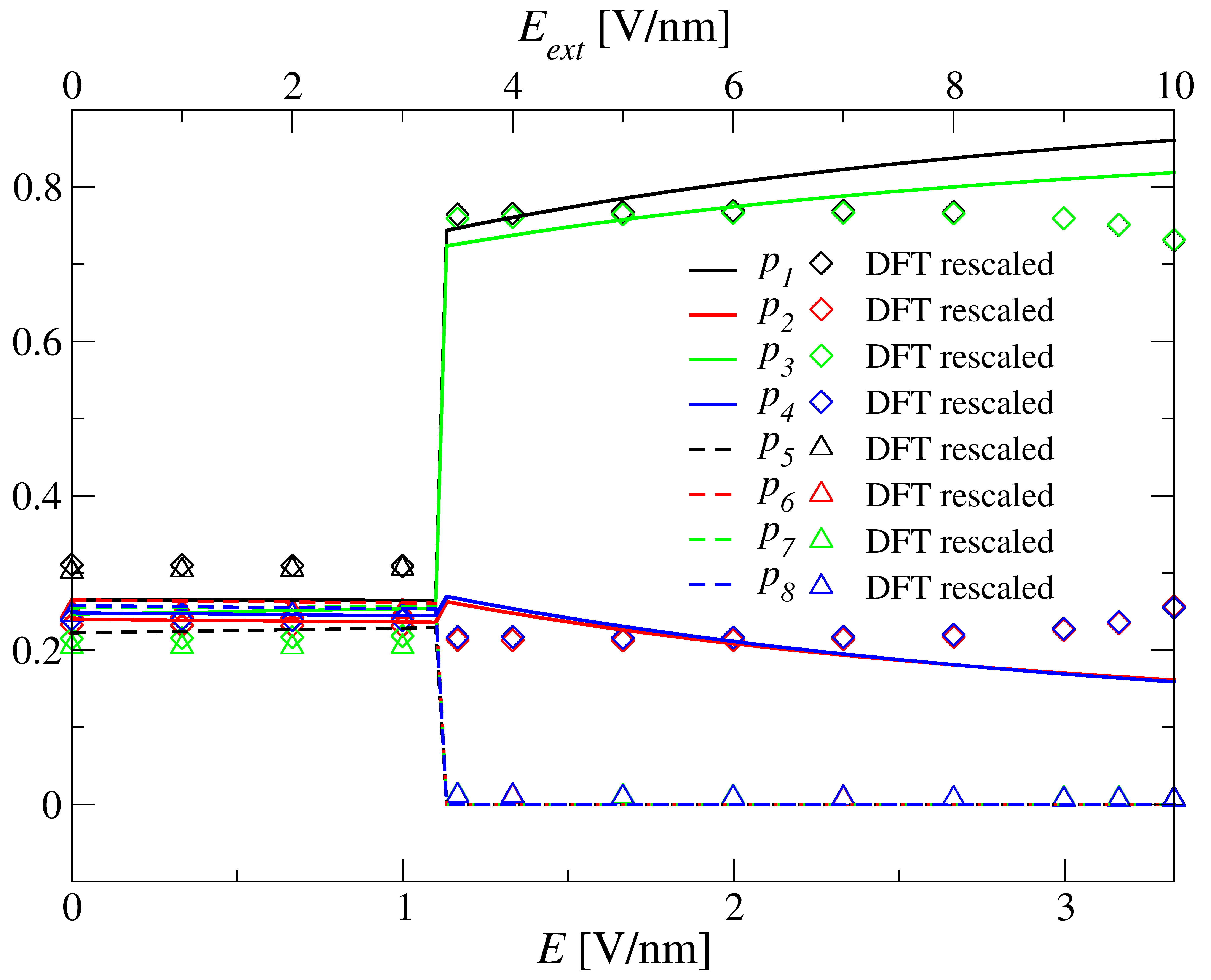}
	\caption{Electric field dependence of probability $p_i$ to find an unpaired electron at a given vanadium site in the ES of molecule {\bf I} at $T=2$ K and $B$=0. Diamonds and triangles stand for the rescaled DFT spin density obtained in the external field $E_{ext}$ (upper scale). Electric field is applied along sites $1$ and $5$. }
	\label{loc15t2go}
\end{figure}

As can be seen in Fig. \ref{loc15t2go} at certain value of the electric field $E_c=1.1$ V/nm ($E_{ext}$=3.3 V/nm) the localization of the electrons suddenly changes. At low field the electrons are almost uniformly distributed in both ES, like for $E=0$, whereas for larger fields in one of the ES (broken lines and triangles in Fig. \ref{loc15t2go}) there are practically no electrons and in the other one two electrons are localized mostly at sites 1 and 3. The situation in the occupied ES is similar to that in the molecule {\bf II} without electric field, but here the occupation of sites 1 and 3 is a bit smaller and that of sites 2 and 4 a bit larger than in molecule {\bf II}, which is due to the smaller value of parameter $\epsilon_2$.

\begin{figure}[h]
	\centering	
	\includegraphics[height=6.5cm]{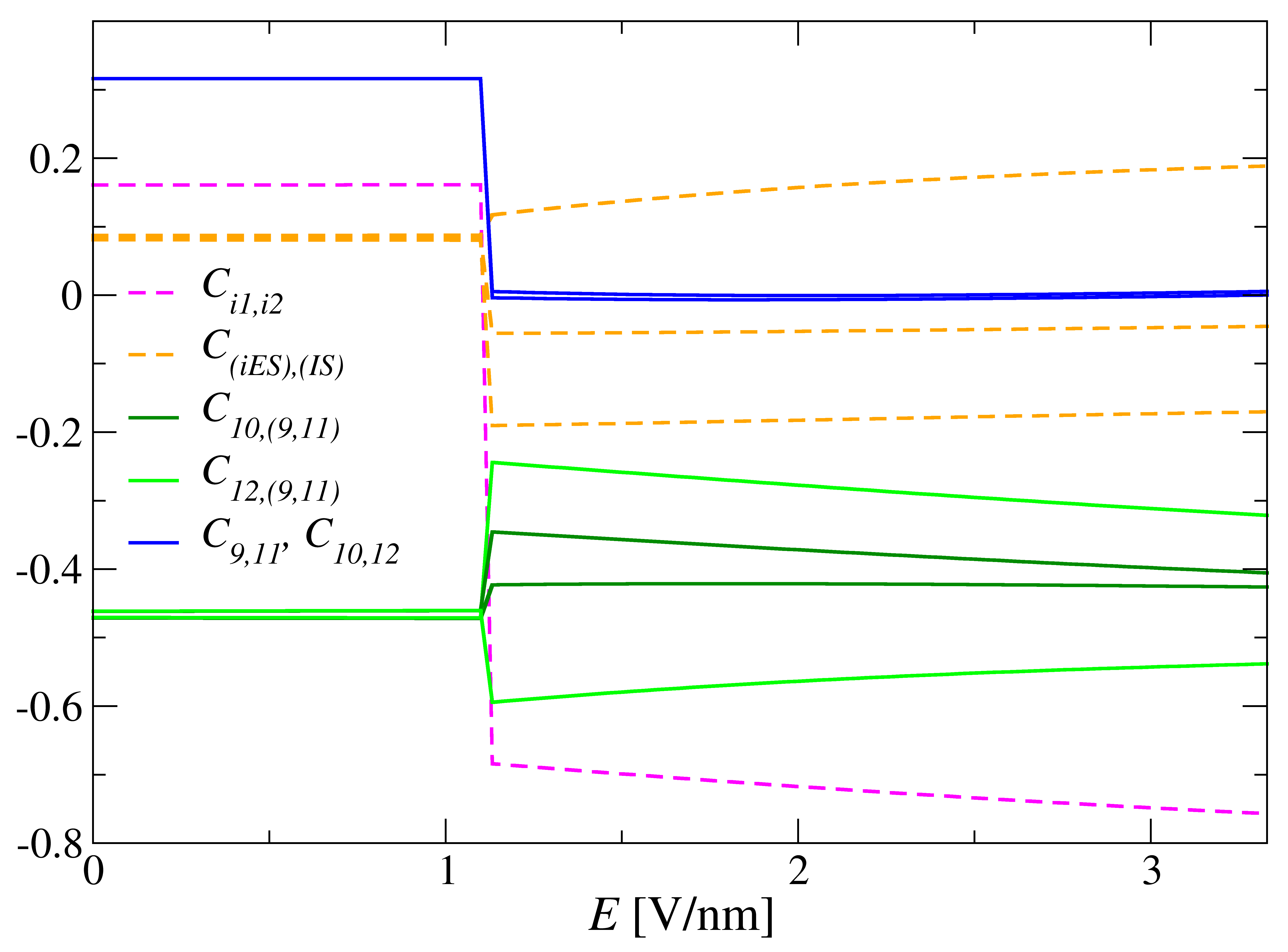}
	\caption{Magnetic correlations for molecule {\bf I} versus electric field applied along sites 1 and 5 at $T=2$ K and $B=0$.}
	\label{corn15-got2-B00}
\end{figure}

This change in electron localization is reflected in magnetic correlations at zero magnetic field (Fig. \ref{corn15-got2-B00}) and in magnetization in non-zero magnetic field. In both cases a sharp change is observed at $E=E_c$. For electric fields smaller than $E_c$ the correlations are the same like for $E=0$ (see Fig. \ref{coref0}). For larger fields correlations between itinerant electrons become strongly antiferromagnetic and their correlations with the electrons in the IS split into three branches: the positive branch corresponds to the interaction with the nn, whereas the two negative ones correspond to the interaction with the more distant electrons. Correlations between the nn in the internal square split into four branches and those between the nnn vanish for higher fields.
The global and local magnetizations at $B=2$ T remain the same as in the molecule without electric field up to $E=E_c$ and then almost vanish for higher electric fields. 

The ground state $S=0$ does not change when the electric field increases, but the excited states do. For $E<E_c$ the two excited energy levels $S=1$ and $S=2$ lie very close to the ground state, like for $B=0$. For larger electric fields the $S=1$ state is about 10 K above the ground state followed by next $S=0$ state lying some 15 K above the ground state. It should be emphasized that for $E<E_c$ both the ground and the excited states correspond to the 1-4-1 electron distribution, whereas for $E>E_c$ the ground and two excited states belong to 2-4-0 distribution. Thus, at $E=E_c$ a crossing of two $S=0$ energy levels corresponding to two different electron distributions takes place. (Similar crossings in a proper energy scale are visualized for molecule II in fig. \ref{fener} in Appendix \ref{apham}.)
Though here the magnetic ground state is not switched by the electric field the magnetic behavior of the molecule {\bf I} is changed abruptly et $E_c$ due to the change of the energy gap between the ground state and the lowest excited state. Thus in practice one can also observe here a kind of spin crossover.

\subsubsection{Molecule II}
In the case of molecule {\bf II} the application of the electric field perpendicular to the ES leads to two abrupt changes in electron distribution from 2-4-2 to 3-4-1 (1-4-3) and then to 4-4-0 (0-4-4). 
The parameters of the Hamiltonian obtained from fitting the $E=0$ data are kept intact and new parameters $\Delta E_o$ and $\Delta E_o^{'}$ are added to account for the change of orbital energy. $\Delta E_o$ is added to the energy of states with distribution 3-4-1 or 1-4-3 and $\Delta E_o^{'}$ to the energy of states with distribution 4-4-0 or 0-4-4. Their values ($\Delta E_o=-21100$ K, $\Delta E_o^{'}=-64882$ K) are fitted so that the relation between the local field $E$ and the external field $E_{ext}$ is the same like for the electric field applied parallel to the ES. Parameter $J_3$ could not be determined by fitting the experimental data. It is assumed that $J_3=100$ K. Its value is important only for 4-4-0 (0-4-4) distribution, but has no influence on the electron distribution. 

\begin{figure}[h]
	\centering	
	\includegraphics[height=7cm]{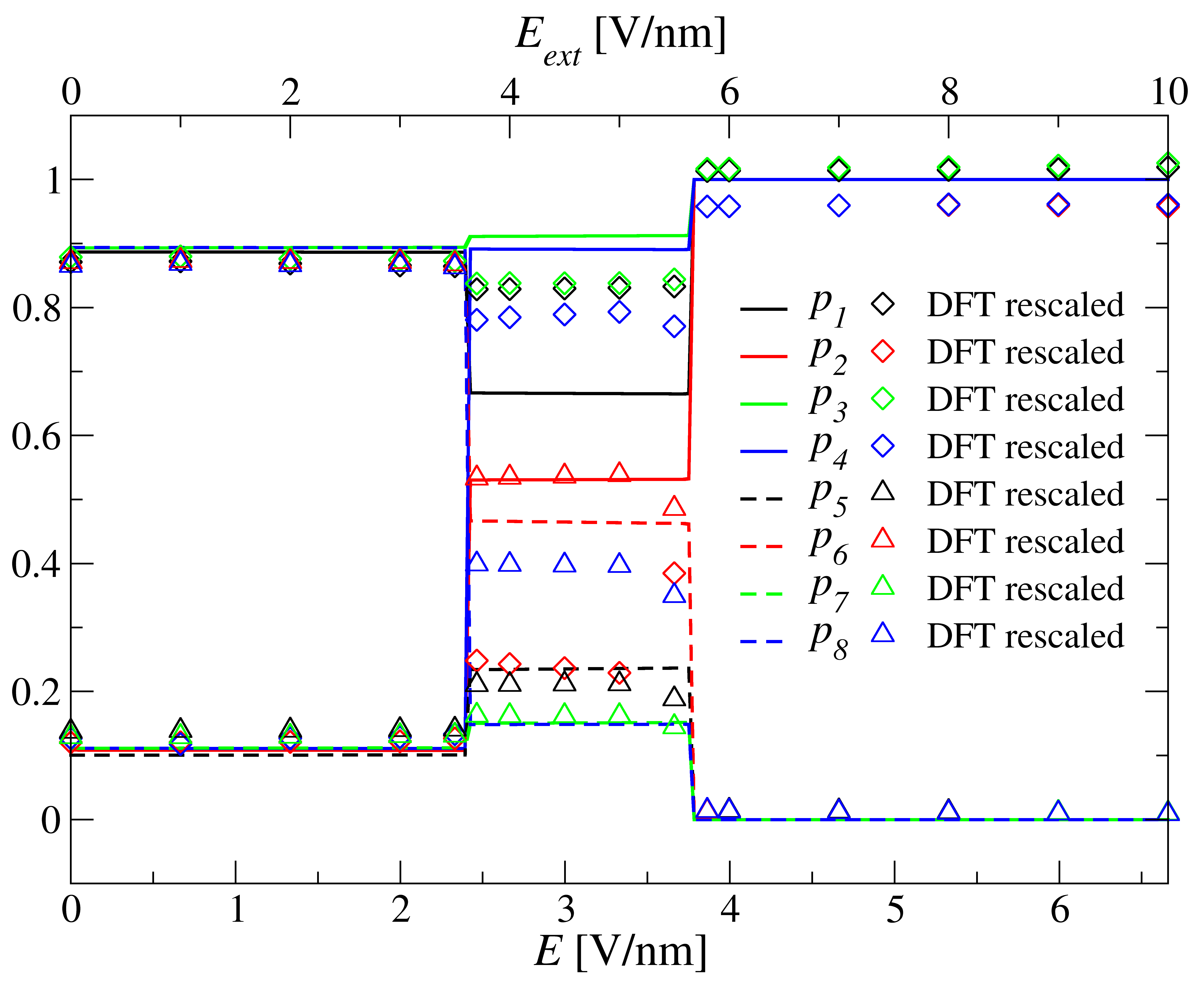}
	\caption{Electric field dependence of probability $p_i$ to find an unpaired electron at a given vanadium site in the ES of molecule {\bf II} at $T=2$ K and $B$=0. Solid and broken lines mark quantities corresponding to different ES. Diamonds and triangles stand for the rescaled DFT spin density obtained in the external field $E_{ext}$ (upper scale). Electric field is applied along sites $1$ and $5$.}
	\label{loc15-B0II}
\end{figure}

As can be seen in Fig. \ref{loc15-B0II} the electron distribution changes at fields $E_{c1}=2.4$ V/nm and $E_{c2}=3.8$ V/nm. Distribution 2-4-2 is conserved below $E_{c1}$, distribution 3-4-1 is present between fields $E_{c1}$ and $E_{c2}$ and distribution 4-4-0 appears for fields larger than $E_{c2}$. There is a very good agreement between the Hamiltonian calculations and DFT for fields smaller than $E_{c1}$ ($E_{ext}<3.6$ V/nm) or larger than $E_{c2}$  ($E_{ext}>5.7$ V/nm). For $E$ between $E_{c1}$ and $E_{c2}$ the agreement is worse. In this region corresponding to the $E_{ext}$ between $3.6$ and $5.7$ V/nm the convergence of DFT calculations is very slow. This can be due to the fact that there are many states with similar energy but different electron distribution which usually leads to very slow convergence. Nevertheless the type of distribution (3-4-1) in this region perfectly matches the Hamiltonian calculations.

\begin{figure}[h]
	\centering	
	\includegraphics[height=6.5cm]{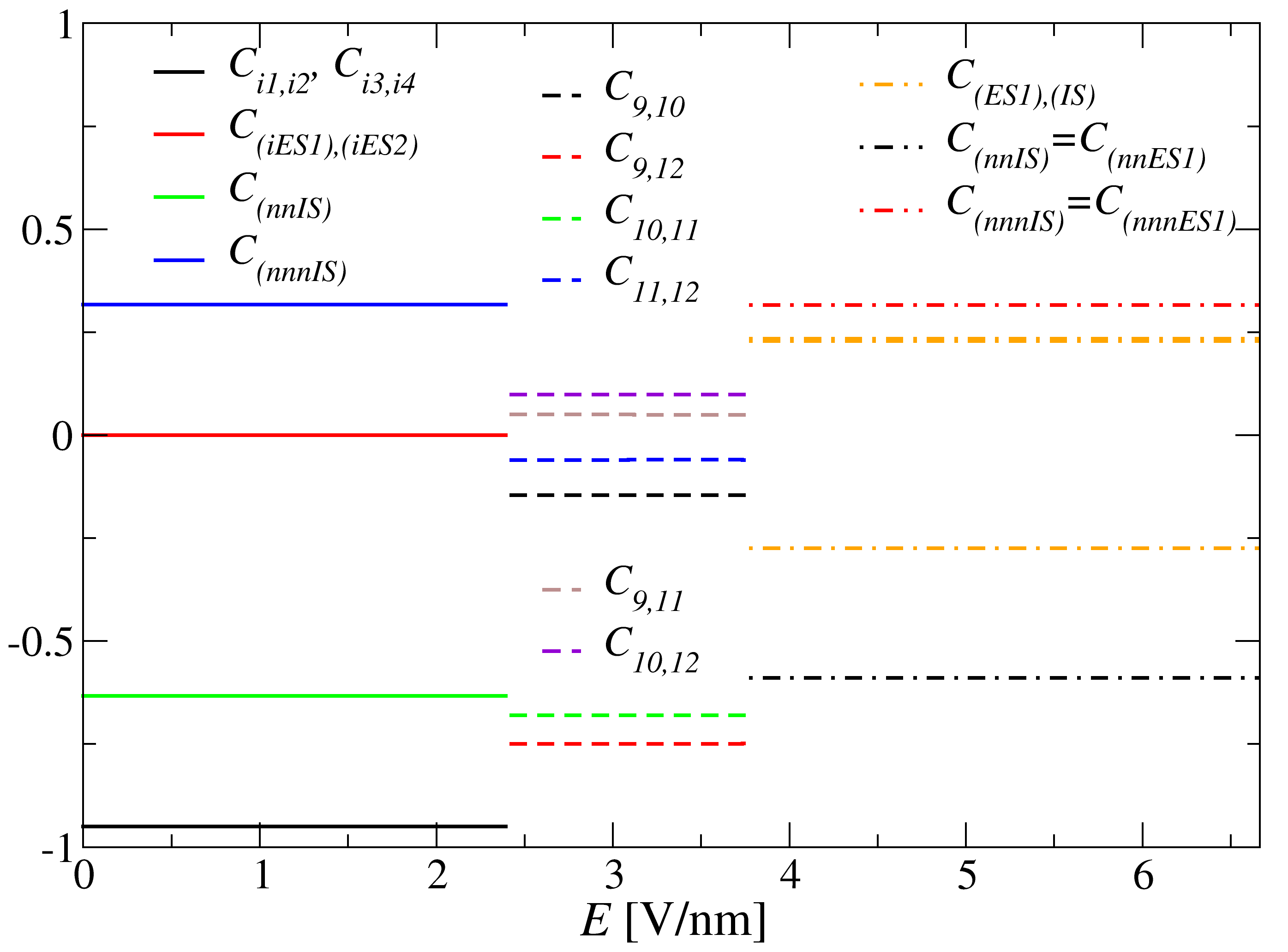}
	\caption{Magnetic correlations for molecule {\bf II} versus electric field applied along sites 1 and 5 at $T=2$ K and $B=0$.}
	\label{corn15-got2-B00II}
\end{figure}

Of course the variation in electron distribution should lead to changes in spin correlations and magnetisations. The electric field dependence of correlations at zero magnetic field is presented in Fig. \ref{corn15-got2-B00II}. 

Below $E_{c1}$ Correlations are the same like for $E=0$: no correlations between itinerant electrons from different ES and strong antiferromagnetic correlations between itinerant electrons within the ES. There are negative correlations between the nn and positive between nnn within the IS. The correlations between itinerant electrons and the electrons in the IS vanish. 

For electric field between $E_{c1}$ and $E_{c2}$ the correlations with the itinerant electrons are not well defined since in the ES1 there are three indistinguishable electrons and it is hard to choose a well defined couple. Therefore they are not shown. 
Within the IS one can observe weak positive correlations between nnn ($C_{9,11}$, $C_{10,12}$). The correlations between nn are weak negative ($C_{9,10}$, $C_{11,12}$) or strong negative ($C_{9,12}$, $C_{10,11}$). The variation in nn correlation is probably induced by uneven distribution of itinerant electrons.

For electric field larger than $E_{c2}$ all the itinerant electrons are localized at the ES1 and correlations between them are the same like between electrons in the IS: negative between nn and positive between nnn. The correlations between electrons from the ES1 and the IS have three values: positive between nearest neighbors (like 1 and 9) and those on diagonal (like 1 and 11) and negative between nnn (like 1 and 10).
Different (but positive) values of $J_3$ (but not too large - see section \ref{fitting}) induce only some changes in values of correlations for fields $E>E_{c2}$ with their general character (sign and relation to other values) conserved.

\begin{figure}[h]
	\centering	
	\includegraphics[height=6cm]{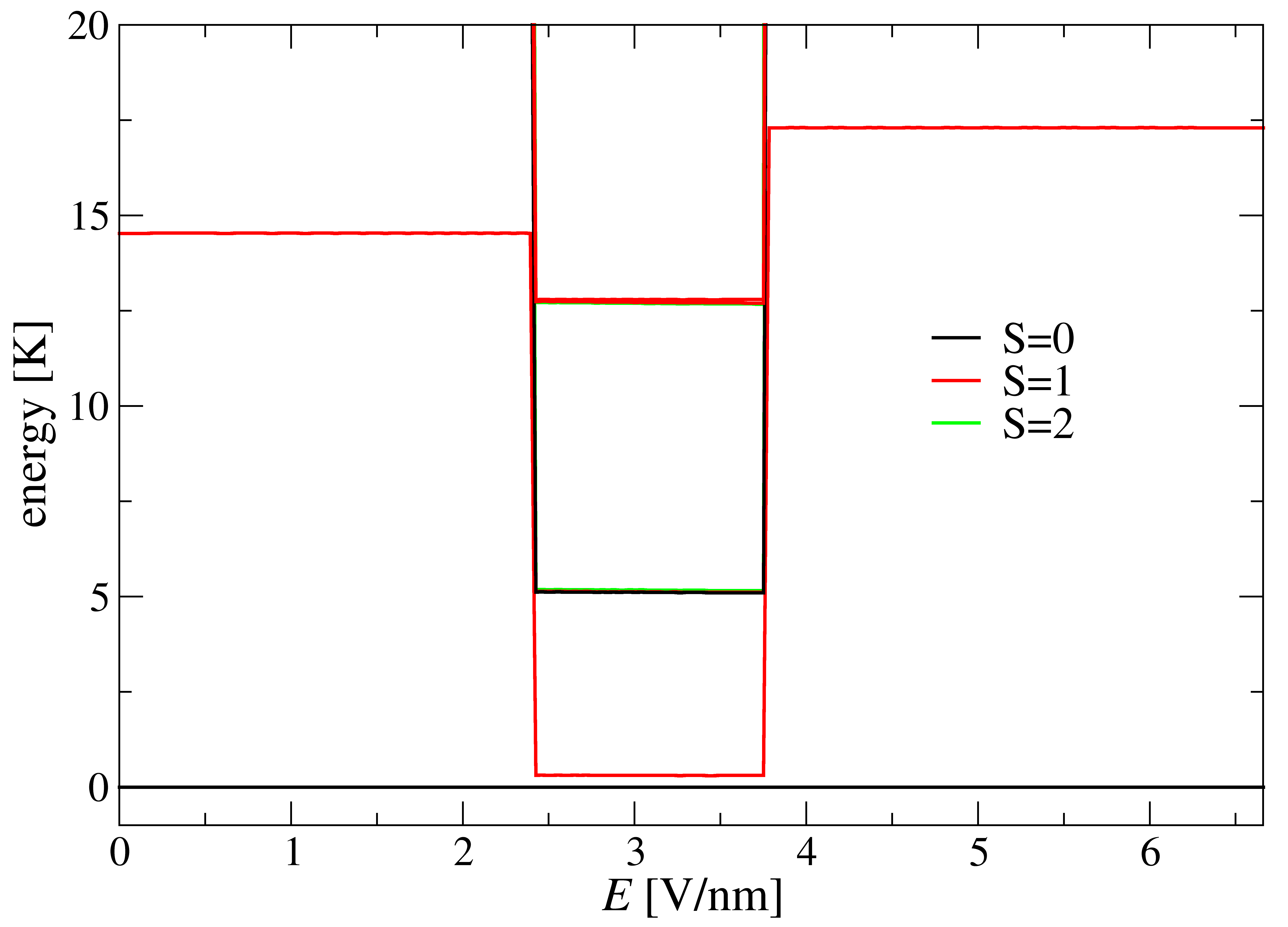}
	\caption{The ground state and a couple of the lowest excited states of molecule {\bf II} as a function of electric field $E$ along sites $1$ and $5$ at $B$=0. The energy is rescaled so that the ground state has energy equal to $0$.}
	\label{groundsn15II}
\end{figure}

The global and local magnetizations vanish at zero magnetic field and for $B=2$ T they are nonzero and constant only if the electric field is between $E_{c1}$ and $E_{c2}$. If $J_3$ is equal zero the nonzero magnetizations appear also for electric field larger than $E_{c2}$. This behavior can be explained by the change of the lowest energy levels visualized in Fig. \ref{groundsn15II}.

Here again like in the case of molecule {\bf I} the magnetic ground state does not change with the electric field, but only the gap between the ground state and the lowest excited states does. The crossings of the lowest energy levels (in a different energy scale) corresponding to different electron distributions are shown in Fig. \ref{fener} in Appendix \ref{apham}. As a result abrupt changes of magnetic behavior can be observed as the electric field is increased. Thus, one can see here a kind of spin crossover behavior as the system is switching from the low spin state to the high spin state and again to the low spin state as the electric field rises.

\section{Conclusions}
Two V$_{12}$ molecules analyzed in this article change their magnetic  properties when exposed to external electric field.  It has been demonstrated, using two complementary theoretical methods: DFT and effective Hamiltonian calculations, that the magnetoelectric effect in these molecules is induced mostly by the relocation of the itinerant electrons leading to their stronger localization. Though, in both molecules for each field direction a clear displacement of the itinerant electrons was observed, only in some circumstances the magnetic behavior was modified.

Generally two patterns of electron displacement have been observed: 1. gradual, for electric field  applied parallel to the ES, corresponding to smooth change of spin densities and 2. abrupt,  for field applied perpendicular to the ES, corresponding to a sudden change of spin densities. In the second situation itinerant electrons were effectively transferred between two ES at some critical values of the electric field.

In molecule {\bf I} gradual displacement leads to the change of the ground state resulting in the change of the magnetic behavior assuming the form of a crossover from the high to the low spin state as the electric field increases. In molecule {\bf II}, on the contrary, gradual displacement generates no magnetic effect. The DFT results suggest that for electric fields larger than $7$ V/nm for {\bf I} and $6$ V/nm for {\bf II} some of spin density is moved out of vanadium cores, which leads to the electron distribution that cannot be modeled by Hamiltonian (\ref{ham}).

For electric field applied perpendicular to the ES both molecules behave in a similar way, as one (for {\bf I}) or two (for {\bf II}) electron transfers between the ES are induced by increasing electric field, leading to abrupt changes of magnetic properties. Here the magnetic ground state remains the same $S=0$ and the change in magnetism is generated by the change of the energy gap between the ground state and the lowest excited state. Thus, as the electric field increases one can observe a crossover from the high to the low spin state in molecule {\bf I} and from the low to the high and again to the low spin state in  molecule {\bf II}. 

In principle the experimental verification of the obtained results seems to be straightforward, as it is enough for instance to measure magnetic susceptibility in the electric field \cite{Wang2023} (see Fig. \ref{chimag42}). However a high electric field required for the effect to appear makes this kind of measurements impossible for macroscopic samples. Yet, electric field of order V/nm can be generated by an STM tip \cite{Moors2023}. Thus, the experimental verification should be done rather in microscopic single molecule or single layer setup. The experimental details should be determined by taking into account possible oxidation/reduction processes that can take place in such a setup (see for instance \cite{Lehmann2007,Huang2025}).

The lowest value of the external electric field at which the magnetic state is changed is equal to $3.3$ V/nm, which is a substantial improvement compared to [GeV$_{14}$O$_{40}$]$^{-8}$ anion \cite{Cardona-Serra2015}  for which the critical external electric field was estimated to be larger than $11$ V/nm. This result indicates that maybe, by proper chemical engineering, it would be possible to lower the critical electric field even more. For this reason one could consider how to modify the molecule to change the values of such parameters like $t$ or $\epsilon_2$ responsible for the delocalization of the electrons, or how to modify orbital energy in the IS. 

\begin{figure}[h]
	\centering	
	\includegraphics[height=6cm]{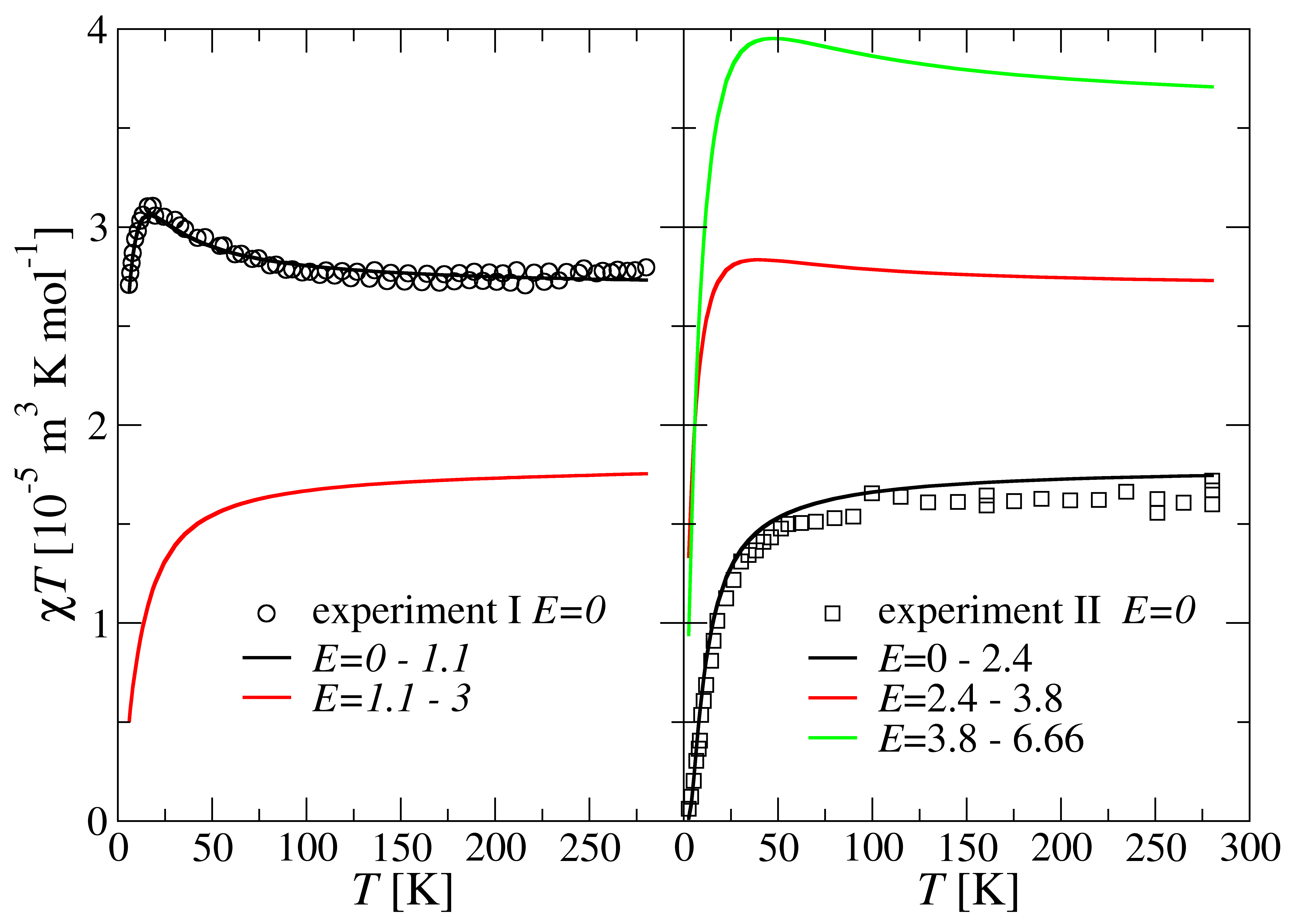}
	\caption{Temperature dependence of $\chi T$ for molecule {\bf I} (left panel) and {\bf II} (right panel) in electric field $E$ in V/nm applied along sites 1 and 5. Symbols stand for experimental results \cite{Gatteschi1993} for molecule {\bf I} (circles) and {\bf II} (squares) at $E=0$.}
	\label{chi15}
\end{figure}

It is worth noting that the change in magnetic properties demonstrated in molecule {\bf I} is perceptible up to temperature $100$ K for the electric field parallel to the ES (see Fig. \ref{chimag42}), whereas for the electric field applied perpendicular to the ES the effect is detectable in both molecules for any temperature between $0$ and $300$ K (see Fig. \ref{chi15}). However the price to be payed for the resistance of the magnetoelectric effect to high temperature is the high value of the electric field triggering the spin crossover. I expect that lowering of the critical electric field will lead to larger dependence on temperature. It has been demonstrated \cite{Bosch-Serrano2012a} that in a mixed valence dimer with double exchange a strong vibronic coupling can lead to smaller predicted values of critical electric fields. Thus, in principle, the introduction of vibronic coupling could give smaller predicted values for critical electric fields. 

The change in electron spin correlations indicates that maybe also quantum entanglement of the electron spins can be manipulated with the electric field which can open the way to quantum computing applications. Further works on these molecules will be devoted to the verification of this scenario and to the determination of other possible electric field effects, like for instance electro-caloric effect. 

\begin{acknowledgments}
Part of the model Hamiltonian calculations have been carried out in Poznańskie Centrum Superkomputerowo-Sieciowe in Poland. The author would like to thank Xavier L\'{o}pez for help with the ADF package.
\end{acknowledgments}

\appendix
\section{DFT calculations \label{apdft}}
The DFT calculations have been carried out with the  ADF2024.1 and ADF2025.1 packages from SCM \cite{adf1,adf2}. 
The Slater-type basis orbitals with triple-$\zeta+$polarisation (TZP) \cite{Lenthe2003} with frozen cores \cite{Baerends1973} (1s for C and O, 1s-3d for As and 1s-3p for V) have been used. Scalar relativistic effects have been taken into account using the Zeroth Order Regular Approximation (ZORA) \cite{Lenthe1993,Lenthe1994,Lenthe1999}. No geometric constrains were applied.
 
The structure obtained by XRD (CCDC - 1260088) have been optimized separately for both molecules  assuming high spin configuration, unrestricted spin and using  BP86 functional \cite{Swart2003}.

For single point calculations B3LYP functional \cite{Stephens1994} have been used with TZ2P basis set, Grimme's D3 dispersion correction \cite{Grimme2010} and conductor-like screening model (COSMO) \cite{Pye1999} with acetonitrile as a solvent. Atomic spin densities have been calculated with Bader QTAIM method \cite{Rodriguez2009,Rodriguez2013}.

To confirm that the frozen core approximation does not influence significantly the calculated atomic spin densities the all-electron calculations have been carried out in a couple of representative points. Only for molecule {\bf II} in the electric field applied along sites 1 and 5 and for $3.6$~V/nm~$<E_{ext}<$~$5.7$~V/nm the frozen core approximation for some vanadium ions gave quite different values of spin densities than the all-electron approach. Yet, in this region the DFT calculations are very slowly converging and do not agree very well with the model Hamiltonian results. However the general character of the results is conserved.

\section{Model Hamiltonian calculations \label{apham}}
The experimental data of molar susceptibility \cite{Gatteschi1993} have been corrected for diamagnetism \cite{bain2008} of a compound, counter ions and solvant [NHEt$_3$]$_2$[NH$_2$Me$_2$][V$_{12}$As$_8$0$_{40}$(HC0$_2$)]$\cdot$ 2H$_2$O for molecule {\bf I} and Na$_5$[V$_{12}$As$_8$0$_{40}$(HC0$_2$)]$\cdot$ 18H$_2$O for molecule {\bf II} \cite{Muller1991}. 

Molar susceptibility and magnetization have been calculated by numerical diagonalization of the Hamiltonian using libraries  BLAS, LAPACK and ScaLAPACK.
For fitting the experimental data an evolutionary algorithm was used to minimize the fitting error:

\begin{equation}
	\Delta=\sqrt{\frac{1}{N}\sum_{i=1}^N\left(\frac{x_i^e-x_i^t}{x_i^t}\right)^2},\label{delta}
\end{equation}
where $N$ stands for the number of experimental points and $x_i^e$ and $x_i^t$ stand for experimental and theoretical results respectively. $\Delta$ can be also expressed in percent by multiplying equation (\ref{delta}) by 100. For the best fit $\Delta=1 \%$ for molecule {\bf I} (the upper panel in Fig. \ref{fit}) and $\Delta=17\%$ for molecule {\bf II} (the lower panel in Fig. \ref{fit}).

\begin{figure}[h]
	\includegraphics*[width=0.42\textwidth]{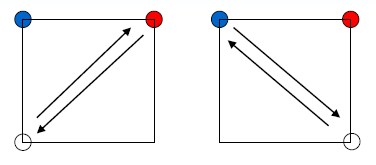}\\
	\includegraphics*[width=0.42\textwidth]{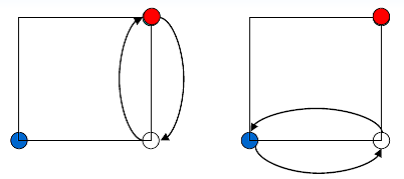}
	
	\caption{Examples of exchange transfer in the ES. upper panel: interactions engaging only nearest neighbor distributions $\epsilon_{ijk}=\epsilon_1$, lower panel: interactions engaging nearest neighbor and diagonal distributions $\epsilon_{ijk}=\epsilon_2$}\label{e1}
\end{figure}

In Figure \ref{e1} the examples of two different types of correlated hoping in the ES are presented. Red and blue circles stand for initial positions of the two spins with opposite polarization. The arrows indicate the hoping and the empty circle stands for the final position of one of the electrons. The arrows are pointing in two directions as the process is reversible.

\begin{figure}[h]
	\includegraphics*[width=0.5\textwidth]{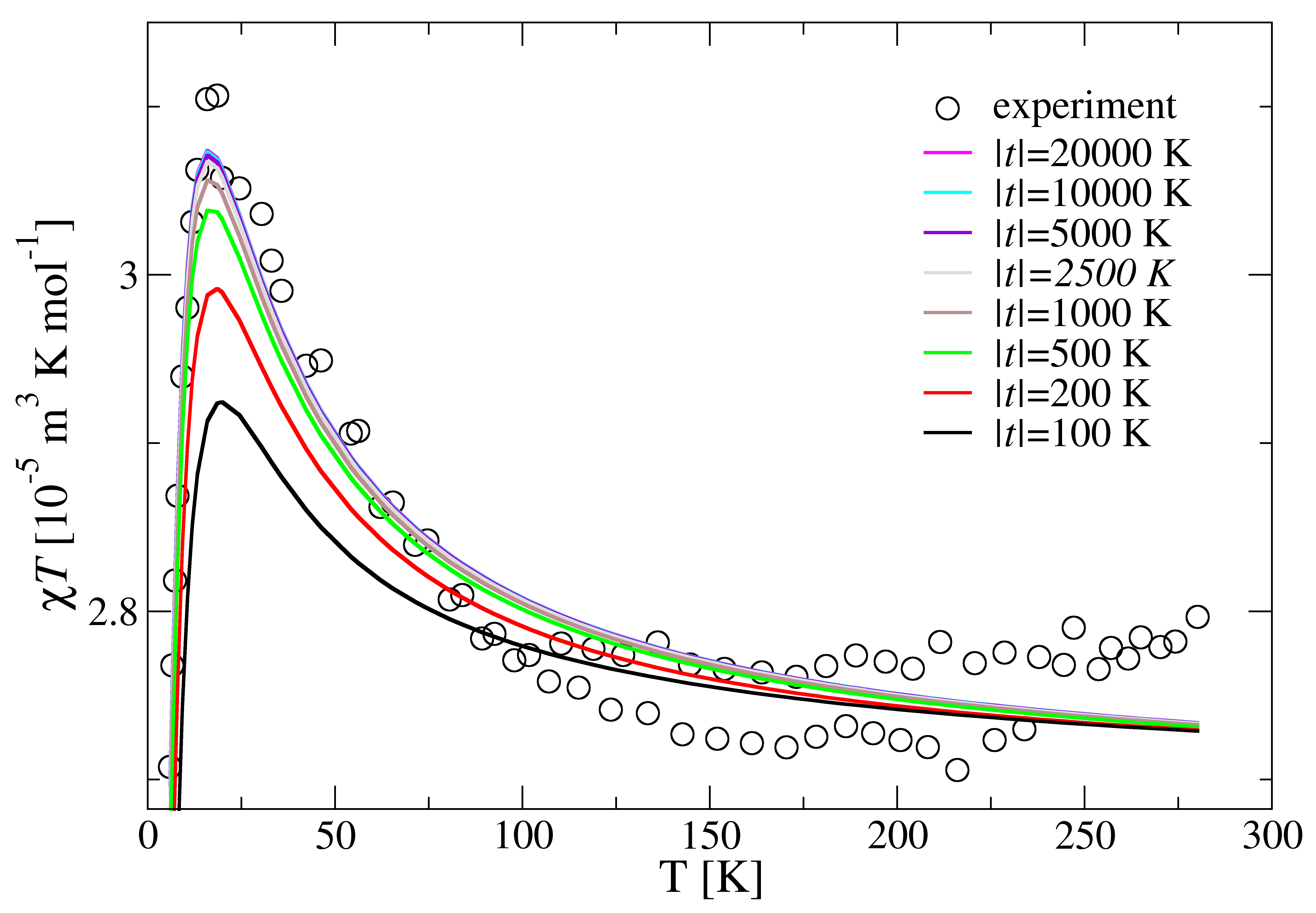}
	\caption{Temperature dependence of $\chi T$ for molecule {\bf I} as a function of parameter $t$ for $E=0$ and $B=0.1$ T}\label{tdep}
\end{figure}

In Fig. \ref{tdep} the $t$ dependence of molar susceptibility $\chi$ calculated with the t-J Hamiltonian (Eq. \ref{ham}) is presented. There is no perceptible difference for $|t|>2000$ K. The remaining parameters are the same like for the best fit.

\begin{figure}[h]
	\includegraphics*[width=0.5\textwidth]{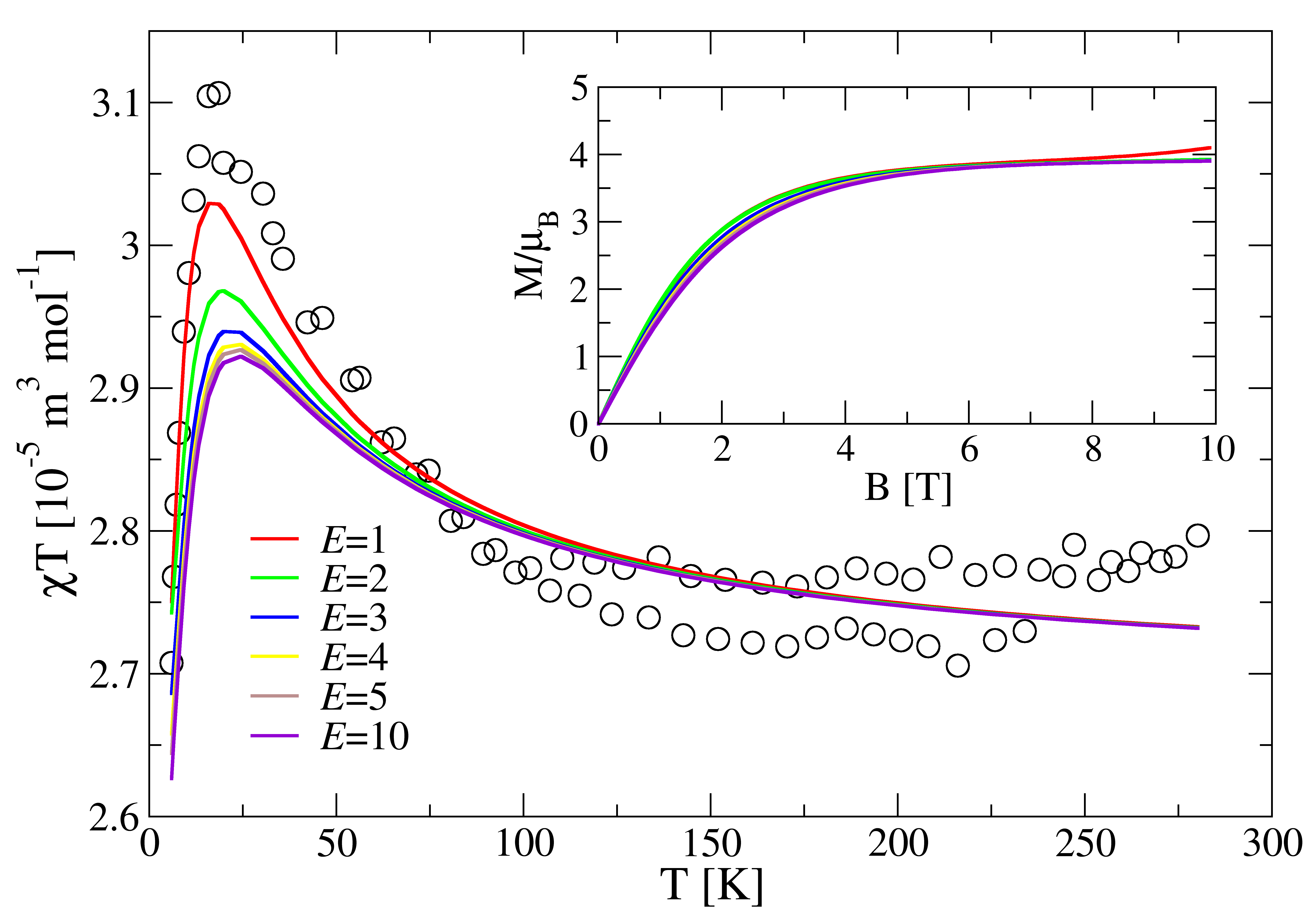}
	\caption{Temperature dependence of $\chi T$ and magnetic field dependence of magnetization $M$ for molecule {\bf I} in different electric fields $E$ [V/nm] applied along sites 1 and 2. Empty circles stand for experimental results \cite{Gatteschi1993} at field $E=0$.}\label{chimagn12}
\end{figure}

\begin{figure}[h]
	\includegraphics*[width=0.5\textwidth]{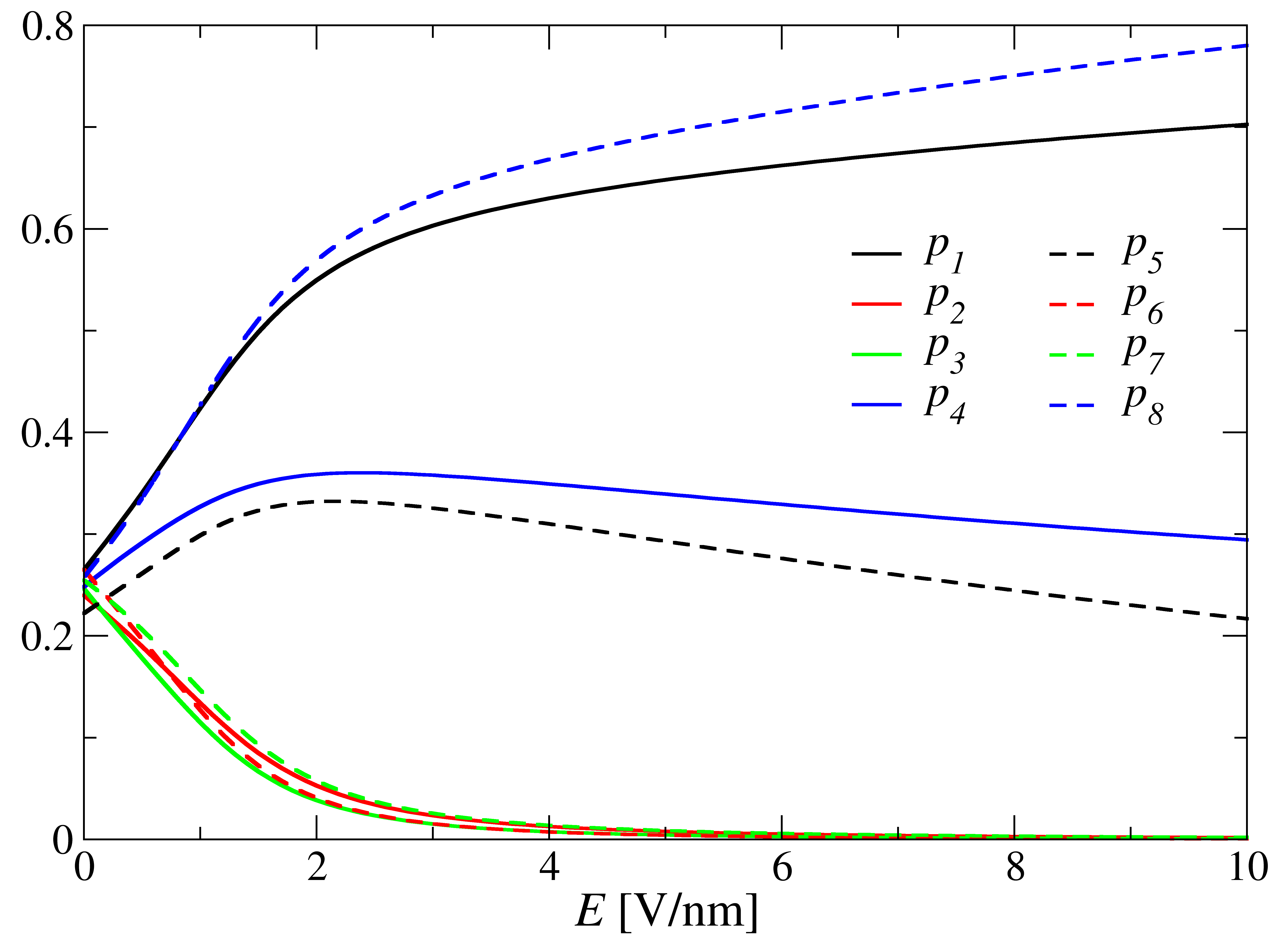}
	\caption{Electric field dependence of probability to find an unpaired electron at a given vanadium site in the ES of molecule {\bf I} at $T=2$ K and $B$=0. Electric field is applied along sites $1$ and $2$.}\label{loc12I}
\end{figure}

\begin{figure}[h]
	\includegraphics*[width=0.5\textwidth]{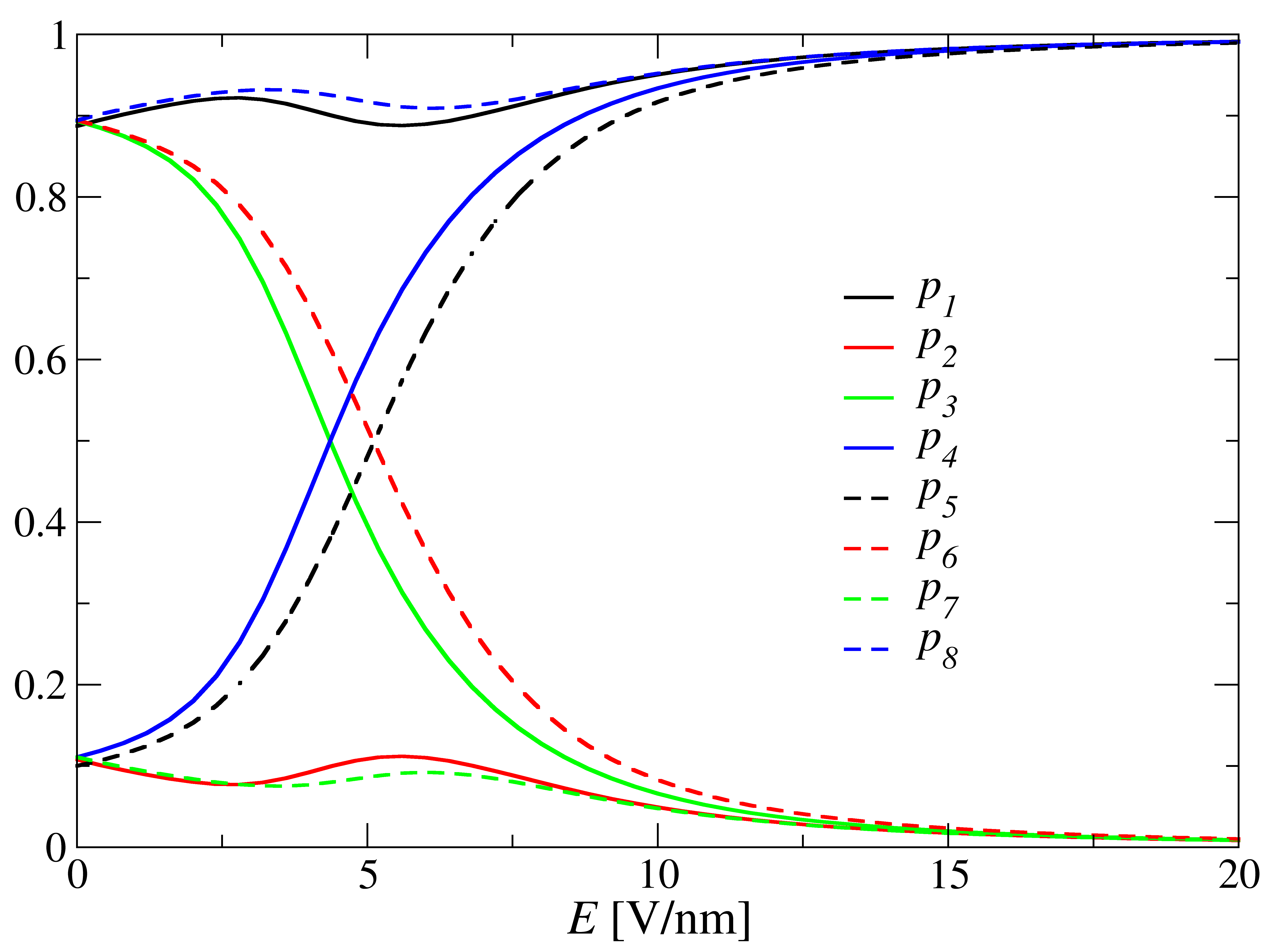}
	\caption{Electric field dependence of probability to find an unpaired electron at a given vanadium site in the ES of molecule {\bf II} at $T=2$ K and $B$=0. Electric field is applied along sites $1$ and $2$.}\label{loc12II}
\end{figure}

In figures \ref{chimagn12}, \ref{loc12I} and \ref{loc12II} various results  for the electric field aligned along sites 1 and 2 are presented.

\begin{figure}[h]
	\includegraphics*[width=0.5\textwidth]{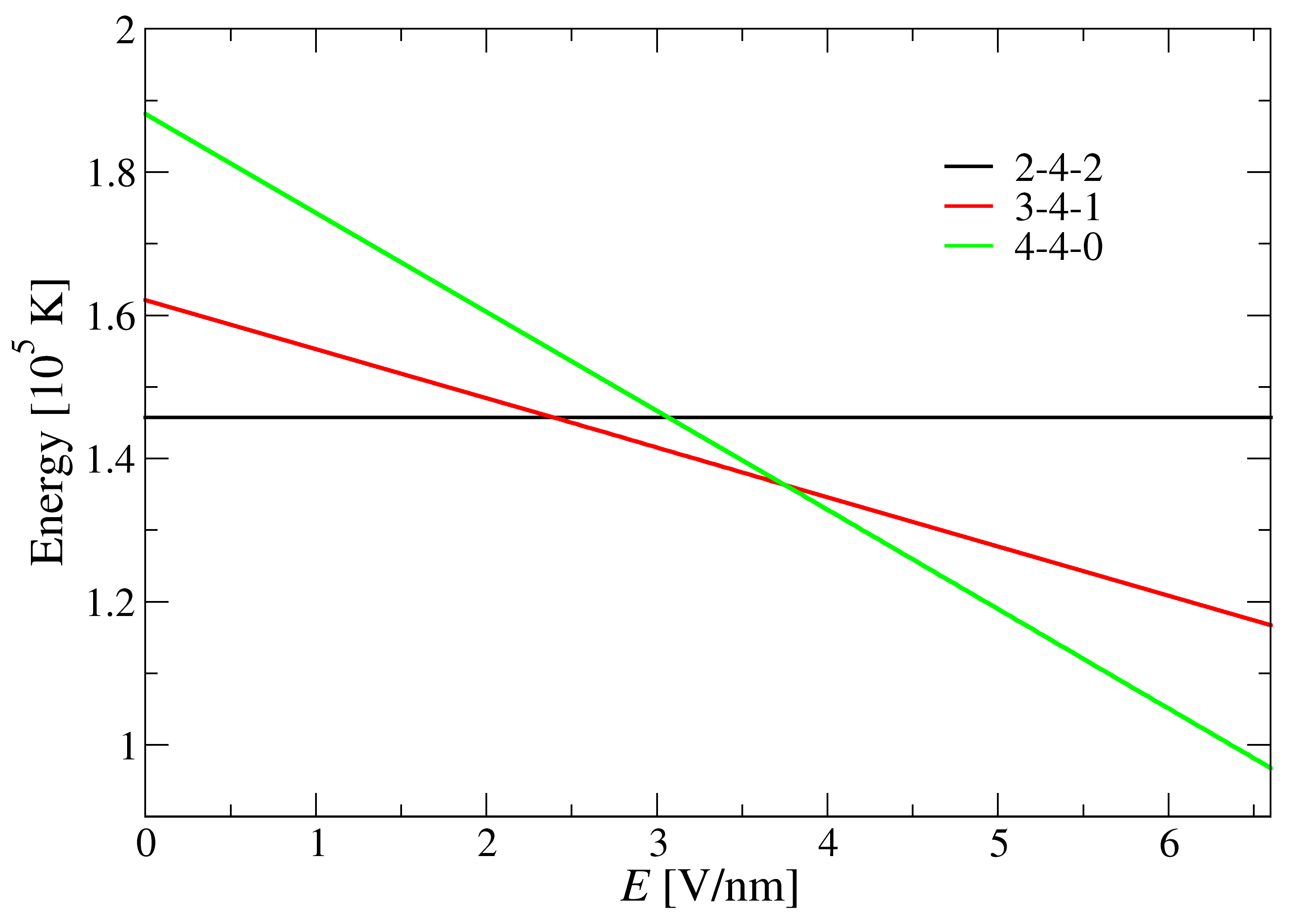}
	\caption{$S=0$ ground states for molecule {\bf II} in different ranges of electric field, corresponding to three different electron distributions (see the legend). The crossings of theses energy levels correspond to electron transfer between the ES as the electric field applied along sites $1$ and $5$ is changed.
	}\label{fener}
\end{figure}
\clearpage
\bibliography{v12lit}

\end{document}